\documentclass[aps,prd,twocolumn,superscriptaddress]{revtex4-2}
\usepackage{amsmath,amsfonts,amssymb,graphicx,graphics,color,hyperref}
\usepackage[latin9]{inputenc}
\usepackage{pifont}
\usepackage{comment}
\usepackage{subcaption}
\usepackage{epsfig,epstopdf}
\usepackage{bm}
\usepackage{enumerate}
\usepackage{float}
\newcommand{\C}{\frac{K+\mu^2_I}{K}}

\begin{document}

\author{Sergio L. Cacciatori}
\email{sergio.cacciatori@uninsubria.it}
\affiliation{Department of Science and High Technology, Universit\`a dell'Insubria, Via Valleggio 11, IT-22100 Como, Italy}
\affiliation{INFN sezione di Milano, via Celoria 16, IT-20133 Milano, Italy, and Como Lake centre for AstroPhysics (CLAP), DiSAT, Universit\`a dell'Insubria, via Valleggio 11, 22100 Como, Italy}

\author{Fabrizio Canfora}
\email{fabrizio.canfora@uss.cl}
\affiliation{Centro de Estudios Cient\'\i ficos (CECs), Avenida Arturo Prat 514, Valdivia, Chile.}
\affiliation{Facultad de Ingenier\'ia, Universidad San Sebasti\'an, sede Valdivia, General Lagos 1163, Valdivia 5110693, Chile.}

\author{Evangelo Delgado}
\email{edelgado2019@udec.cl}
\affiliation{Departamento de F\'isica, Universidad de Concepci\'on, Casilla, 160-C, Concepcion, Chile.}

\author{Federica~Muscolino}
\email{federica.muscolino@gmail.com}
\affiliation{Dipartimento di Matematica e Applicazioni, Universit\'a di Milano Bicocca, via
Roberto Cozzi 55, 20125 Milano, Italy}
\affiliation{Gruppo Nazionale di Fisica Matematica, InDAM, Piazzale Aldo Moro 5, 00185 Rome, IT}

\author{Luigi Rosa}
\email{rosa@na.infn.it}
\affiliation{Dipartimento di Matematica e Applicazioni ``R. Caccioppoli''
Universit\`a di Napoli Federico II, Complesso Universitario di Monte S. Angelo,
Via Cintia Edificio 5A, 80126 Naples, Italy}
\affiliation{INFN, Sezione di Napoli, Complesso Universitario di Monte S. Angelo,
Via Cintia Edificio 6, 80126 Naples, Italy}

\title{Thermodynamics of magnetized BPS baryonic layers and the effects of the isospin chemical potential}

\begin{abstract}
    Through the Hamilton-Jacobi equation of classical mechanics, BPS magnetized baryonic layers (possessing both baryonic charge and magnetic flux) have been constructed in the gauged non-linear sigma model (G-NLSM) minimally coupled to Maxwell theory. This is one of the most relevant effective theories for Quantum Chromodynamics (QCD) in the strongly interacting low-energy limit, which also accounts for electromagnetic interactions. Since the topological charge that naturally appears on the right-hand side of the BPS bound is a non-linear function of the baryonic charge, the thermodynamics of these magnetized baryonic layers is highly non-trivial. In this work, using tools from the theory of the Casimir effect, we derive analytical relationships between baryonic charge, topological charge, magnetic flux, and relevant thermodynamic quantities (such as pressure, specific heat, and magnetic susceptibility) for these layers. The critical baryonic chemical potential is identified. Interestingly, the grand canonical partition function can be related to the Riemann zeta function. On the technical side, it is a remarkable result to derive explicit expressions for all these thermodynamic quantities for a strongly interacting magnetized system at finite baryon density. Moreover, thanks to the BPS property of these configurations, we are also able to analytically determine both the equation of state and the speed of sound. The effects of the isospin chemical potential can also be included; specifically, we are able to explicitly construct the BPS bound and the corresponding BPS configurations even when the isospin chemical potential is non-zero. The physical interpretations of our analytical results are discussed. 
\end{abstract}

\maketitle

\section{Introduction}\label{Introduction}

Quantum Chromodynamics (QCD) is the field theory that describes the strong nuclear interaction. One of its unique features is the behavior of its coupling constant under renormalization flow: it increases as the energy decreases, and vice versa. This results in a phase diagram divided into different regimes, each requiring distinct tools or approximations to study specific conditions. In the low-energy limit of QCD and under extreme conditions, neither perturbative methods (due to the increasing coupling constant) nor numerical simulations (due to the sign problem in Lattice QCD) are feasible (see \cite{nambu1961dynamical,rajagopal2001frontier,alford2001crystalline,casalbuoni2004inhomogeneous} and references therein).

Studying QCD under extreme conditions, such as finite baryon density, low temperatures, and intense magnetic fields, is of great physical interest because these conditions occur in neutron stars and heavy-ion collisions \cite{bzdak2020mapping,nagata2022finite,astrakhantsev2021lattice,astrakhantsev2021electromagnetic,brandt2023thermal,busza2018heavy,yagi2005quark,dorso2018phase,dorso2020nucleonic,lopez2021properties,pelicer2023anisotropic,nandi2018transport,yakovlev2015electron}. There is both numerical and phenomenological evidence suggesting that, in this limit, non-homogeneous baryonic condensates exist, forming regular structures known as nuclear pasta phases \cite{ravenhall1983structure,hashimoto1984shape,horowitz2015disordered,berry2016parking,dorso2018phase,schneider2018domains,caplan2018elasticity,nandi2018calculation,lin2020fast,dorso2020nucleonic,pethick2020elastic,lopez2021properties,schuetrumpf2019survey,barros2020fluctuations,acevedo2020warming}. Developing analytical theories to support this evidence is crucial.

In particular, there are very few examples in which the thermodynamics of interacting systems with high baryonic charge confined in a finite spatial volume can be studied analytically. The main goal of the present manuscript is to provide a concrete example of a strongly interacting system with high baryonic charge, where it is possible to study the thermodynamics with analytical tools. As we will see, such a system also has an intrinsic physical relevance.

Interesting results have been obtained in this direction in the low-energy limit of QCD \cite{skyrme1954new,skyrme1955meson,skyrme1958non,skyrme1961non,skyrme1961particle,skyrme1962unified}. In such a limit, baryons arise as solitonic solutions (Skyrmions) in a mesonic field theory. The relations between the low energy limit of QCD and QCD itself have been investigated by several authors \cite{t1993planar, witten1979baryons,callan1984monopole,witten1983current,witten1983global,balachandran1983low,balachandran1984doubly,BALACHANDRAN1985525,ADKINS1983552,manton2004topological,balachandran1991classical}. There is convincing evidence that relates the properties of baryons to Skyrmions (see for instance \cite{ADKINS1983552}). 

In order to describe the hadronic structures arising in neutron stars, in \cite{chen2014exact,canfora2013nonlinear,ayon2016analytic,aviles2017analytic,canfora2018analytic,canfora2019analytic,ayon2020vera,alvarez2017integrability,canfora2018ordered,canfora2019crystals,canfora2020crystals,barsanti2020analytic,canfora2021pion,canfora2021gravitating,canfora2021exact,cacciatori2021analytic,alvarez2020analytic,cacciatori2022cooking,cacciatori2024pearcey} two ansatze have been proposed. The resulting exact analytic solutions correspond to hadronic matter distributed in tubes (or spaghetti states) and layers (lasagna states). However, until very recently, no purely magnetic solution was constructed (as all the available analytic configurations minimally coupled with the Maxwell field possessed both electric and magnetic fields of
the same order). The present study will focus specifically on purely magnetic baryonic layers. 

We will consider the gauged nonlinear sigma model (G-NLSM) in (3+1) dimensions with global SU(2) symmetry. It is worth mentioning here that the reason we consider the NLSM and not the full Skyrme model lies in the fact that the Skyrme term is necessary in order to guarantee the stability of the solution defined \textit{over infinite three-dimensional space} \cite{manton2004topological}. On the other hand, our solutions are defined \textit{over a finite and fixed space volume}, in which case the Derrick theorem does not apply. Using an appropriate ansatz, we will describe magnetized baryonic layers and, through the Hamilton-Jacobi technique, derive a nontrivial BPS bound that will serve as the foundation of our work. 

A novel technique using the Hamilton-Jacobi equation has enabled us to derive a non-trivial BPS bound for the G-NLSM, relating topological charge to baryonic charge in a nonlinear manner \cite{canfora2023magnetized,canfora2024superconducting,canfora2025fractional,canfora2025novel}. To find an analytical relationship between topological charge, baryonic charge, and magnetic flux, we must determine the ``chemical potential'' for the magnetic flux ($I_0$), which emerges as an integration constant when solving the Hamilton-Jacobi equations. Although an exact analytical form for $I_0$ does not exist, we will seek a sufficiently accurate approximation using computational methods. Obtaining these quantities analytically is crucial for constructing the grand canonical partition function of the system, enabling the calculation of free energy, heat capacity, and a deeper understanding of the thermodynamics of the G-NLSM.

As it has already been emphasized, the main goal of the present manuscript is to provide a concrete example of a strongly interacting system with high baryonic charge, where it is possible to study the thermodynamics with analytical tools. Usually, this task is impossible (unless the baryons are assumed to be free, which is clearly not the case in the present context), while, due to the BPS property of the configurations which will be analyzed in the following sections, such an analysis is allowed.

On the other hand, since the shape of the configurations analyzed in the following sections corresponds to magnetized baryonic layers, a natural question is whether our analysis can also be of relevance for the nuclear Lasagna phase. In order to describe the nuclear pasta phase, both the isospin chemical potential and the electromagnetic interactions must be taken into account. 

As far as the electromagnetic interactions are concerned, these will be included through the minimal coupling of the SU(2) valued chiral field with the U(1) gauge field through the covariant derivative. Indeed, this is the reason why we will consider the gauged Non-Linear Sigma Model minimally coupled to Maxwell theory: our exact solutions are self-consistent solutions of the complete set of field equations, which take into account both the backreaction of the electromagnetic fields on the hadronic fields and vice versa.

As far as the effects of the isospin chemical potential are concerned, at first glance, one could think that the situation is very difficult, as the presence of the isospin chemical potential could spoil the BPS property of the baryonic magnetized layers. The reason is that the isospin chemical potential explicitly enters into the field equations, and consequently, it also modifies the solutions. Nevertheless, quite remarkably, the BPS property (which is the basis of our analysis) is not destroyed, but only modified. We will be able to construct explicitly isospin-dependent BPS magnetized baryonic layers. However, for the sake of clarity, it is better to first analyze the system without the isospin chemical potential and then to include its effects. The reason is that in most of the formulas, the effects of the isospin chemical potential manifest themselves by replacing some of the coupling constants with "modified coupling constants" dressed by the effects of the isospin chemical potential itself.

This paper is organized as follows: in Section \ref{G-NLSM}, we present the G-NLSM. Section \ref{MagBPS} is devoted to the ansatz for the construction of magnetized baryonic layers, the corresponding BPS bound, the topological charge and its relation to the magnetic flux, as well as a discussion on the boundary conditions for $H(r)$. In Section \ref{Approx}, we introduce a couple of approximations for the integration constant $I_0$. The grand canonical partition function is constructed in Section \ref{Thermo}, where we also compute several relevant thermodynamical quantities. In Section \ref{MagSusc}, we explore the contribution of an external magnetic field to the energy and compute the magnetic susceptibility. In \ref{EqState&SpeedSound} we analytically calculate the equation of state and the speed of sound thanks to the BPS property of the system. In Section \ref{IsospinChemical}, the effects of the isospin chemical potential will be included, and we will construct the BPS bound and the corresponding BPS configurations in the case in which the isospin chemical potential is non-zero. Section \ref{Conclusion} contains some comments and concluding remarks.

\section{Gauged non-linear sigma model}\label{G-NLSM}
In this section, we introduce the main properties of our model. As already mentioned in the introduction, we first consider the case with the isospin chemical potential $\mu_I=0$. The coupling of the model with $\mu_I\neq 0$ and its effects will be directly introduced in Section \ref{IsospinChemical}.  

In order to describe the low-energy limit of QCD minimally coupled to electromagnetism, let us consider the gauged non-linear sigma model action:
\begin{equation}
    S=-\frac{1}{4}\int d^4 x \left\{K\mathrm{Tr}\{\Sigma^{\mu}\Sigma_{\mu}\}-F_{\mu\nu}F^{\mu\nu}\right\},
\end{equation}
where $K=\frac{f_\pi ^2}{4}$ is the coupling constant of the G-NLSM and $f_\pi$ is the pion decay constant \cite{adkins1983static}. The indices $\mu,\nu=0,1,2,3$ are the usual spacetime indices, where $0$ represents the timelike component. In addition, 
\begin{equation}\label{DefSigmaF}
    \Sigma_\mu = U^{-1}D_\mu U,\quad F_{\mu\nu}=\partial_\mu A_\nu - \partial_\nu A_\mu,
\end{equation}
where $A_\mu$ represents the $U(1)$ gauge field and $U$ is a scalar field with values in $SU(2)$, defined as a map 
\begin{equation}
    U:\mathbb{R}^{3}\rightarrow SU(2).
\end{equation}
In the following sections, the map $U$ will be parametrized using the \textit{generalized Euler angles} \cite{bertini2006euler,cacciatori2017compact,tilma2004generalized} as follows 
\begin{equation}\label{U}
    U=e^{\tau_3 F}e^{\tau_2 H}e^{\tau_3 G}.
\end{equation}
Here $F(x^\mu)$, $H(x^\mu)$ and $G(x^\mu)$ are the three pseudoscalar functions of the spacetime
\begin{align}
    F,G,H:\mathbb{R}^{3+1}\mapsto\mathbb{R}
\end{align}
and $\tau_j = i\sigma_j$, where $\sigma_j$ are the Pauli matrices. As already studied in, for instance, \cite{alvarez2020analytic}, this parameterization leads to configurations where baryons are organized into \textit{layers}.

Going back to the definition \eqref{DefSigmaF}, the covariant derivative is defined as follows
\begin{equation}
    D_\mu U = \partial_\mu U + A_\mu[\tau_3, U].
\end{equation}
Infinitesimal variations of the action give the equations of motion
\begin{align}
    D_\mu \Sigma^\mu&=0,\\
    \partial_\mu F^{\mu\nu} &= J^\nu,
\end{align}
where the conserved current $J^\mu$ reads 
\begin{equation}
    J^\mu = \frac{K}{2}\mathrm{Tr} \left\{ \hat{O}_3 \Sigma^\mu \right\}, \quad \hat{O}_3 = U^{-1}\tau_3 U -\tau_3.
\end{equation}
Furthermore, the energy-momentum tensor for the G-NLSM is
\begin{equation}
    T_{\mu\nu}=-\frac{K}{2}\mathrm{Tr}\left\{\Sigma_\mu \Sigma_\nu - \frac{1}{2}g_{\mu\nu}\Sigma^\alpha \Sigma_\alpha \right\} + \Bar{T}_{\mu\nu},
\end{equation}
where
\begin{equation}
    \Bar{T}_{\mu\nu}=F_{\mu\alpha}{F_{\nu}}^{\alpha}-\frac{1}{4}F_{\alpha\beta}F^{\alpha\beta}g_{\mu\nu}.
\end{equation}

Notably, different solutions of the equations of motion are labeled by a topological invariant (that we are going to call $B$) defined as the index of the map $U$ ($B=\mbox{Index}(U)$). This index takes integer values and is associated with the number of baryons contained in the system (in other words, the \textit{baryonic charge}) \cite{manton2004topological}. It can be explicitly computed as follows
\begin{equation}\label{baryonicCharge}
    B=\frac{1}{24\pi^4}\int \rho_B,
\end{equation}
where the integral is defined over the volume of the system and the  $\rho_B=\rho_{B_1} + \rho_{B_2}$ represents the \textit{baryonic density} (see \cite{cacciatori2022cooking} and references therein), where 
\begin{align}\label{rhoB1}
    \rho_{B_1} &= \varepsilon^{ijk}\mathrm{Tr}\left\{(U^{-1}\partial_i U)(U^{-1}\partial_j U)(U^{-1}\partial_k U)\right\},\\\label{rhoB2}
    \rho_{B_2} &= -3\varepsilon^{ijk}\mathrm{Tr}\left\{ \partial_i \left[ A_j \tau_3 (U^{-1}\partial_k U + \partial_k U U^{-1}) \right]\right\}.
\end{align}

\section{Magnetized BPS baryonic layers}\label{MagBPS}
As discussed above, the choice of an ansatz is crucial in order to obtain analytic solutions to the equations of motion (i.e., the Skyrme equations). In this section, we study an ansatz corresponding to magnetic solutions in the BPS limit.

To this end, let us compute explicitly the quantities introduced in the previous section using the parameterization \eqref{U} and equipping the spacetime with the flat metric
\begin{equation}
    ds^2=-dt^2 + L^2(dx^2+dy^2)+L_{r}^{2} dr^2,
\end{equation}
where $x$, $y$ and $r$ are dimensionless Cartesian coordinates having ranges 
\begin{equation}
    0\leq x \leq \pi, \quad 0\leq y \leq 2\pi, \quad 0\leq r \leq 2\pi.
\end{equation}
The \textit{dimensionality} of the space is enclosed in $L$ and $L_r$. A more accurate analysis on the units of measure and their specific values will be discussed later. This metric describes a box of volume $V=4\pi^3L_r L^2$, where the gauged solitons live. The area of the layer is given by $A=2\pi^2L^2$. It is important to mention that the coordinates $x$ and $y$ are tangential to the layer, while the coordinate $r$ is orthogonal (see \cite{alvarez2020analytic,cacciatori2021analytic,cacciatori2022cooking}). As a consequence, the energy and baryonic density will depend only on the coordinate $r$. 

Let us consider the following static ansatz for the $U$ scalar field and for the $A_\mu$ gauge field
\begin{align}\label{UAnsatz}
    U&=e^{py\tau_3}e^{H(r)\tau_2}e^{px\tau_3},\\ \label{AAnsatz}
    A_\mu &= \left(0,0,\frac{p}{2}-u(r),-\frac{p}{2}+u(r) \right),
\end{align}
where $p$ is an integer and the index $\mu=(0,r,x,y)$. As demonstrated in \cite{alvarez2017integrability,aviles2017analytic,canfora2018analytic}, with the above ansatz, the seven coupled field equations boil down to two
\begin{align}\label{eq1}
    H''+4\left(\frac{L_r}{L}\right)^2 \sin(2H)\left(\left(\frac{p}{2}\right)^2-u^2\right)&=0,\\ \label{eq2}
    u''-4KL_{r}^{2}\sin^2(H)u&=0,
\end{align}
where the prime stands for the derivative with respect to $r$, with the following energy density 
\begin{align}\label{EnDensity}
    T_{00}&=\frac{K}{L^2}\left[p^2\cos^2(H)+4\sin^2(H)u^2\right]\cr
    &\qquad\qquad\qquad\qquad+\frac{K(H')^2}{2L_{r}^{2}}+\frac{(u')^2}{(L_r L)^2}
\end{align}
and baryonic density 
\begin{equation}\label{baryonicDensitySol}
    \rho_B = -12p\frac{d}{dr}[u(1+\cos(2H))].
\end{equation}
Since the latter is a total derivative, the baryonic charge only depends on the boundary conditions of $u$ and $H$, as expected. The choice of these boundary conditions will be discussed later. It is worth emphasizing here that (being the configurations of interest static and purely magnetic) the corresponding field equations are obtained by minimizing the energy, which is equivalent to considering the energy density as a Lagrangian. On the other hand, in the case in which the isospin chemical potential is taken into account, one should minimize the free energy instead. 

\subsection{A novel BPS Bound}

As already discussed in \cite{canfora2023magnetized}, the energy density in equation \eqref{EnDensity} can be rewritten by introducing the Hamilton-Jacobi equation (HJ) 

\begin{align}
    &\frac{L_{r}^{2}}{2K}\left( \frac{\partial W}{\partial H}\right)^2 +  \left( \frac{L_{r}L}{2} \right)^2 \left( \frac{\partial W}{\partial u}\right)^2\cr
    &\qquad\qquad\quad= \frac{K}{L^2}\left[p^2\cos^2(H)+4\sin^2(H)u^2\right],
\end{align}
whose solution is given by
\begin{equation}
    W=\frac{4K^{\frac{3}{2}}}{pL_r}u\cos(H),
\end{equation}
provided the following relation holds:
\begin{equation}
    L=\frac{p}{\sqrt{2K}} \leftrightarrow A=\pi^2\frac{p^2}{K} \leftrightarrow A=\left( \frac{2\pi p}{ef_\pi}\right)^2. \label{p-dependence}
\end{equation}
In this way, it is possible to rewrite the energy density in a BPS style
\begin{align}\label{EnDensityBPS}
    T_{00}&=\frac{K}{2(pL_r)^2}\left[\left(pH'\pm 4K^{\frac{1}{2}}L_r u\sin(H)\right)^2\right.\notag\\
    &\qquad\qquad+\left.4\left(u' \mp pK^{\frac{1}{2}}L_r \cos(H) \right)^2\right] \pm \frac{dW}{dr}.
\end{align}
Through an explicit computation, one finds the following bound

\begin{gather}\label{TopologicalCharge}
    E=\int d^3 x \sqrt{-g} T_{00} = AL_r\int_{0}^{2\pi} dr T_{00} \geq |Q|,
\end{gather}
where
\begin{align}
    Q=\frac{AL_r}{e^2}|W(2\pi)-W(0)|.
\end{align}

The first order BPS equations, which imply the second order field equations, are:
\begin{align}\label{Eq1}
    H'+\frac{4K^{\frac{1}{2}}L_r}{p}u\sin(H)&=0\\\label{Eq2}
    u'-pK^{\frac{1}{2}}L_r\cos(H)&=0.
\end{align}

An observation here is in order. If we divide \eqref{Eq2}  by $p$, the RHS of the equation does not depend on $p$. Therefore, it seems reasonable to assume that the solution $u(r)/p$ will not depend on $p$. For this reason, we are going to define the quantity 
\begin{equation}
    v(r)\equiv\frac{u(r)}{p},
\end{equation}
which is $p$-independent. 

Using the BPS condition \eqref{Eq1} and \eqref{Eq2}, an analytical relationship can be derived between the $SU(2)$ profile $H$ and the gauge field profile $v$. In particular, one obtains the equation

\begin{equation}
    \frac{dH}{dv}=-4v\tan(H),
\end{equation}
with solution
\begin{equation}
H(r)=\arcsin\left[\exp\left(-2v^2(r)-I_0\right)\right]\label{H}
\end{equation}
where $I_0$ is an integration constant. Substituting the above equation into \eqref{Eq2}, the complete set of field equations is reduced to the quadrature:

\begin{equation}\label{uEq}
    v'=K^{\frac{1}{2}}L_r\sqrt{1-\exp\left(-4v^2(r)-2I_0 \right)}
\end{equation}

By integrating the equation \eqref{uEq}, we obtain
\begin{equation}
    2\pi=\frac{1}{\sqrt{K}L_r}\int_{0}^{v(2\pi)} \frac{d\tau}{\sqrt{1-\exp(-4\tau^2 -2I_0)}}, \label{equation}
\end{equation}
where $\tau=v(r)$, and for convenience we choose $v(0)=0$.
For the physical analysis of $v(2\pi)$, it is necessary to calculate the total magnetic flux $\Phi$ in the $y$ direction as follows.

\begin{align}
    \Phi &= LL_r \int drdx F_{rx}= \frac{p^2\pi L_r}{\sqrt{2K}}(v(2\pi)-v(0)),
\end{align}  
which implies that
\begin{align}\label{Phiu}
    v(2\pi) = \frac{\sqrt{2K}}{p^2\pi L_r}\Phi.
\end{align}

\subsection{Topological charge and magnetic field}

Let us now consider the baryonic charge \eqref{baryonicCharge} with the baryonic density \eqref{baryonicDensitySol}. Using the solution of $H$ defined in \eqref{H}, it takes the explicit form 

\begin{align}
    |B|=4\frac{\sqrt{2K}\Phi}{\pi L_r}\left(1-\exp\left(-\frac{8K\Phi^2}{p^4 (\pi L_r)^2}-2I_0 \right) \right).
\end{align}

\begin{figure}
    \centering
    \includegraphics[scale=0.5]{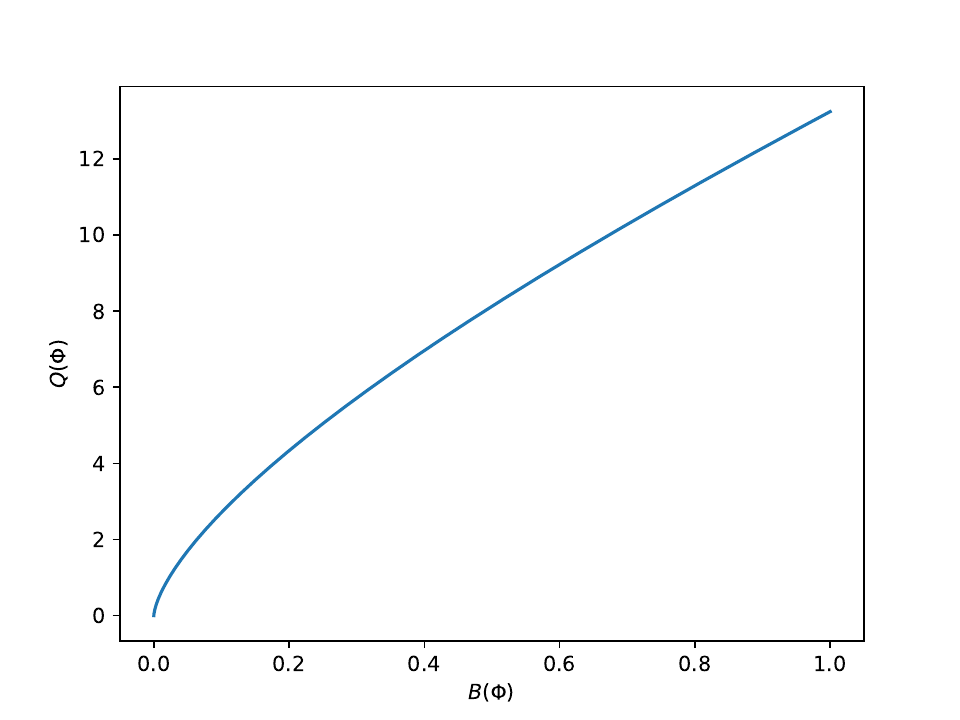}
    \caption{Topological charge as a function of the baryon charge}
    \label{fig:QvsB}
\end{figure}

It is interesting to compare the above expression for the baryonic charge with the topological charge obtained in \eqref{TopologicalCharge}. Indeed, using the expressions of $Q$ \eqref{TopologicalCharge}, $H$ \eqref{H} and $\Phi$ \eqref{Phiu}, we get
\begin{align}
    Q&=\frac{\pi^2p^2L_r}{K}[W(2\pi)-W(0)]\cr
    &=\frac{4\sqrt{2}K\pi\Phi}{L_r}\sqrt{1-\exp\left(-\frac{8K\Phi^2}{p^4 (\pi L_r)^2}-2I_0 \right)}.
\end{align}
From here, it is possible to obtain the relation
\begin{align}
    \frac{Q}{B}=\frac{\pi^2 \sqrt{K}}{\sqrt{1-\exp\left(-\frac{8K\Phi^2}{p^4 (\pi L_r)^2 }-2I_0\right)}}
\end{align}
In this way, we can plot the topological charge as a function of the baryon charge (see Fig. \ref{fig:QvsB}).

As we can see, unlike non-interacting systems, there is a nonlinear dependency between the baryonic charge and the topological charge. However, as both the baryonic and topological charges increase, they enter a linear regime, just as expected.

\subsubsection{On the boundary conditions on $H(r)$ and the meaning of the baryonic charge} \label{Sec:BC}
The definition of the boundary conditions on $H(r)$ deserves a deeper discussion. From the physical point of view, the baryonic density $\rho _{B}$ defined in Eqs. \eqref{baryonicCharge}, \eqref{rhoB1} and \eqref{rhoB2} has to be interpreted physically and concretely as the \textit{baryonic density}: namely, a density with the property that its integral is the baryonic charge. Thus, one could think that the only reasonable boundary conditions are the ones where the baryonic charge is an integer (and this is certainly a reasonable viewpoint). On the other hand, in situations in which the baryonic charge is very large (as in neutron stars) it is also possible to consider the baryonic charge as  a continuous quantity. Indeed, variations of $\pm 1$ baryons are very small compared to a total baryonic charge of the order of $10^{10}$. Under these conditions, baryons can be considered as continuous variations of the mesonic field and the baryonic charge can be described as a continuous variable. In this subsection, we will show that in our case both viewpoints are reasonable and that the second one (where the baryonic charge is considered as a continuous variable) has some technical advantages (see also Section \ref{Sec:ContinuousB}). 

The boundary conditions on the functions appearing at the exponents of equation \eqref{U} are generally defined in such a way that the map $U$ describes a closed manifold, isomorphic to an $S^3$ (see, for instance, \cite{manton2004topological} or \cite{cacciatori2022cooking}). Nevertheless, the introduction of the magnetic field defined by the ansatz \eqref{AAnsatz} and the definition of the BPS equations \eqref{eq1} and \eqref{eq2} compromise the closure of the manifold. In particular, the map $U$ introduced in \eqref{UAnsatz} is isomorphic to an $S^3$ without two points. A simple way to see this is the following. First of all, let us observe that, in order to get a closed manifold, one should impose the following boundary conditions on the exponent of \eqref{UAnsatz}
\begin{equation}\label{Closenss1}
    0\leq y\leq 2\pi,\qquad 0\leq x\leq \pi.
\end{equation}
Furthermore, $H$ should be a continuous function on ${0\leq r\leq 2\pi}$ such that (see, for instance, \cite{alvarez2020analytic,cacciatori2022cooking,bertini2006euler,cacciatori2017compact,tilma2004generalized} for further details)
\begin{equation}\nonumber
    0<H(r)<\frac\pi 2\quad\mbox{or}\quad\frac\pi 2<H(r)<\pi \quad\mbox{for}\quad0< r< 2\pi,
\end{equation}
with boundary conditions
\begin{equation}\label{Closenss21}
    H(0)=\frac\pi 2\quad\mbox{and}\quad H(2\pi)=0
\end{equation}
for the first case, or
\begin{equation}\label{Closenss22}
    H(0)=\frac\pi 2\quad\mbox{and}\quad H(2\pi)=\pi
\end{equation}
for the second case (notice that $\sin(H(r))$ decreases, due to \eqref{H}). The condition \eqref{Closenss1} is automatically satisfied by the definition of the space in which the system is defined. On the other hand, the BPS equations \eqref{eq1} and \eqref{eq2} lead to the solution for $H(r)$ outlined in \eqref{H}. It is straightforward to observe from here that $\sin{H(r)}\neq 0$ for each $r$, thus avoiding the \textit{ending points} defined in \eqref{Closenss21} or \eqref{Closenss22}. The condition $H(0)=\pi /2$ can be obtained when $I_0=0$, but in this case the relation \eqref{equation} implies that $v(0)=v(2\pi)=0$, thus leading to $H(r)=\pi /2$ for each $r$. One can deduce that $H(r)$ does not span the whole interval necessary to close the manifold. This, clearly, also affects the baryonic charge. Indeed, $B$ only depends on the boundary conditions of $v(r)$ and $H(r)$, as follows from \eqref{baryonicCharge} and \eqref{baryonicDensitySol}. For closed manifolds, $B$ takes integer values. In our case, the boundary conditions of $v(r)$ and $H(r)$ are linked together using the equations \eqref{H} and \eqref{equation}. In particular, it is possible to define $I_0$ in terms of $v(2\pi)$ through equation \eqref{equation} and  $v(2\pi)$ in terms of $H(2\pi)$ through equation \eqref{H} (notice that, when we fix $v(0)=0$, then $I_0$ only depends on the choice of $H(0)$). Therefore, one can write the value of $B$ in terms of $H(2\pi)$, $v(2\pi)$ or $I_0$. For instance, we can define everything in terms of $I_0$, which now can assume arbitrary values, since it is not constrained by the \textit{closeness conditions} of the manifold. This way, the baryonic charge $B$ varies continuously with $I_0$.

It is worth mentioning here that the solutions to the equations \eqref{Eq1} and \eqref{Eq2} describe systems with $E=Q$, where $Q>0$, unless $I_0=0$, in which case we have the trivial condition $Q=0$. The parameter $I_0$ can be used to define a continuous deformation of the solutions with $Q>0$ to solutions with $Q=0$. In this sense, the system is not \textit{topologically stable}. Nevertheless, once all the boundary conditions are fixed, the solutions have finite energy for $I_0\neq 0$.

Interestingly, when $I_0\rightarrow\infty$, then $H(r)\rightarrow 0$. In this case, the map $U$ does not wrap around $S^3$ at all. The contribution to the baryonic charge is given entirely by the presence of the magnetic field (in particular, it derives from the volume integral of \eqref{rhoB2}, since the volume integral of \eqref{rhoB1} is zero). In this situation, the quantity $Q$ defined in \eqref{TopologicalCharge} only differs from $B$ by a multiplicative constant. Explicitly, 
\begin{equation}\nonumber
    B=8p^2\pi\sqrt{K}L_r\qquad\mbox{and}\qquad Q=\pi^2\sqrt{K}B.
\end{equation}
Therefore, the obtained solution is characterized by $E= Q>0$, where the positivity of the energy is guaranteed only by the magnetic field contribution.

Now, two questions are in order. Are the solutions obtained from \eqref{Eq1} and \eqref{Eq2} stable for fixed values of $B$? What is the interpretation of the quantity $B$ in this paradigm? 

In order to answer these questions, we need to observe that both $B$ and $Q$ only depend on the boundary conditions. Once the latter are fixed, $B$ and $Q$ are invariant under the symmetries of the system. Then, for each sector characterized by $B$ fixed, also $Q$ is fixed and the energy associated with solutions to the Skyrme equations \eqref{eq1} and \eqref{eq2} has a finite minimum given by equation \eqref{TopologicalCharge}. This minimum is reached by the solutions to the BPS equations \eqref{Eq1} and \eqref{Eq2}. Furthermore, we give $B$ its original interpretation of \textit{baryonic charge}. From a physical point of view, the fact that it can assume continuous values could be interpreted as due to a quantum correction that screens the baryonic charge. This interpretation would deserve a deeper investigation at a more fundamental level, but this is out of the scope of this current paper.

\section{Approximation}\label{Approx}

To analytically obtain the baryonic charge $B$ and the topological charge $Q$ as functions of the magnetic flux, it is essential to determine the integration constant $I_0$ analytically. For this reason, we define the function $F$ as

\begin{equation}\label{FVero}
    F(\Phi, I_0)= \int_{0}^{\frac{\sqrt{2K}}{p^2\pi L_r}\Phi} \frac{d\tau}{\sqrt{1-\exp(-4\tau^2 -2I_0)}}
\end{equation}

Unfortunately, solving the integral in an elementary form is impossible. However, standard techniques from the theory of the Casimir effect (see \cite{bordag2009advances} and references therein) based on the expansion of the integral in terms of Bessel functions provide an excellent analytic approximation. 

\begin{figure*}

    \centering
    \begin{subfigure}{0.4\textwidth}
	    \includegraphics[width=\textwidth]{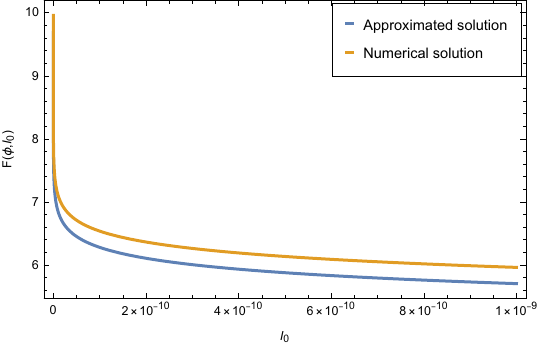}
        \caption{\label{Fig:FApproxSmall}}
    \end{subfigure}\hspace{.3cm}
    \begin{subfigure}{0.4\textwidth}
        \includegraphics[width=\textwidth]{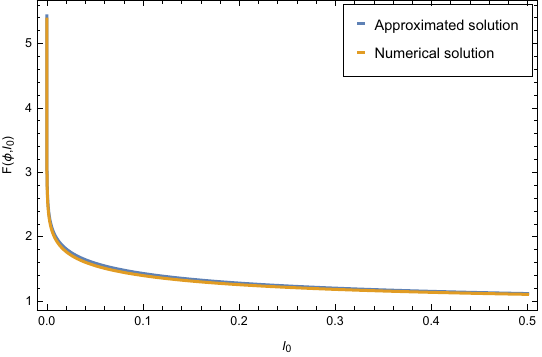}
        \caption{\label{Fig:FApproxBig}}
    \end{subfigure}
    
    \caption{Analytical approximation numerical integration for equation \eqref{FVero}, where ${\sqrt{2K}\Phi}/({p^2\pi L_r})=1$  (which corresponds to $\Phi\approx10^{5}\ \mbox{fm}^2$. See Section \ref{UnitsOfMeasures} for a discussion on the units of measure). The graph \eqref{Fig:FApproxSmall} represents the approximation for small $I_0$, which becomes quite accurate when $I_0\rightarrow 0$. The graph \eqref{Fig:FApproxBig} is the approximation valid for $I_0>0$ big enough.}
    \label{fig:comparation}
\end{figure*}

Therefore, for $I_0>0$ we propose the following analytical approximation for \eqref{FVero}, which, as shown in the graph of Fig. \ref{fig:comparation}, is quite close to the numerical solution of the integral.

\begin{equation}\label{ApproxI0Big}
    F(\Phi,I_0) \approx \frac{\sqrt{2K}}{p^2\pi L_r}\Phi-\frac{1}{4}\ln\left( 1-e^{-2I_0}\right).
\end{equation}

Nonetheless, it fails for $I_0\rightarrow 0$. Indeed, the approximated expression \eqref{ApproxI0Big} diverges for very small values of $I_0$. In this limit, we use the expression
\begin{align}\label{ApproxI0Small}
 F(\Phi,I_0)&\approx\frac 14 \log \left( 1+\frac {8K^2\Phi^2}{p^4\pi^2L_r^2I_0}\right.\notag\\ 
 &\qquad+\left.\sqrt {\frac {8K^2\Phi^2}{p^4\pi^2L_r^2I_0}\Big( 2+\frac {8K^2\Phi^2}{p^4\pi^2L_r^2I_0} \Big)} \right).
\end{align}

The explicit computation of these approximations has been reported in Appendix \ref{App:Approximation}.

\begin{figure}[htb!]
    \includegraphics[scale=.83]{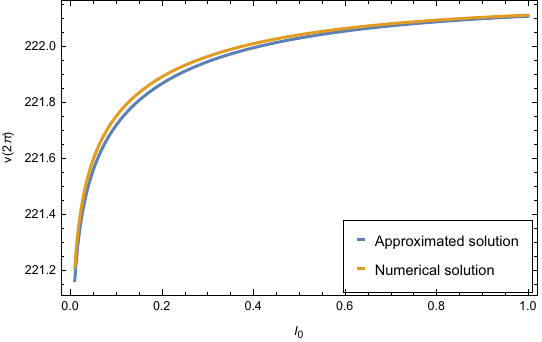}
    	
	\caption{In this graph, it is compared the analytical expression of $v(2\pi)$ given in \eqref{ApproxSol} and the evaluation obtained from a numerical solution of $F(\Phi,I_0) = 2\pi L_r \sqrt{K}$, where $F$ is expressed in \eqref{FVero}. The constant parameters have been fixed to $K=2\ \text{fm}^{-2}$, $L_r =25\ \text{fm}$.}
	\label{ut/p}
\end{figure}

From \eqref{ApproxI0Big} and \eqref{ApproxI0Small}, and given that $F(\Phi,I_0) = 2\pi L_r \sqrt{K}$, one explicitly determine $v(2\pi)$ in terms of the $I_0$. Namely,
\begin{widetext}
\begin{equation}\label{ApproxSol}
	v(2\pi)(I_0)\approx\left\{\begin{array}{l r}
	      \left[2\pi\sqrt{K}L_r+\frac{1}{4}\ln\left(1-e^{-2I_0}\right)\right]\qquad & \mbox{For }I_0\geq\varepsilon,\\
	      \sqrt {\frac {I_0}2} \sinh (4\pi \sqrt K L_r) & \mbox{For }I_0<\varepsilon,
	\end{array}\right.
\end{equation}
\end{widetext}
where $\varepsilon$ is a positive, real quantity that minimizes the error of the approximation with respect to the original relation $F(\Phi,I_0) = 2\pi L_r \sqrt{K}$; it must be determined numerically (in our case, it is of the order of $I_0\approx10^{-12}$). In Fig. \ref{ut/p} , it is represented the comparison between the function \eqref{ApproxSol} and $v(2\pi)$ obtained through a numerical solution of $F(\Phi,I_0) = 2\pi L_r \sqrt{K}$.
\\

Through the equation \eqref{ApproxSol}, we can obtain analytically the dependence of $I_0$ in terms of the boundary conditions $v(2\pi)$: 
\begin{widetext}
\begin{equation}\label{I0Approx}
	I_0(v(2\pi))\approx\left\{\begin{array}{l r}
	      -\frac{1}{2}\log\left[1-\exp\left(4v(2\pi)-8\pi\sqrt{K}L_r\right)\right]\qquad & \mbox{For }v(2\pi)\geq\tilde\varepsilon,\\
	      v^2(2\pi)\frac{2}{\sinh^2 (4\pi \sqrt K L_r)} & \mbox{For }v(2\pi)<\tilde\varepsilon,
	\end{array}\right.
\end{equation}
\end{widetext}
where $\tilde\varepsilon$ has the same role as $\varepsilon$ in \eqref{ApproxSol}. This approximation can be used in order to define an analytical expression for the quantities $B$ and $E$ in terms of $v(2\pi)$, which, in the general case, have the form
\begin{align}\label{BApprox}
	B&=4p^2 v(2\pi)\left(1-e^{-4v^2(2\pi)-2I_0}\right),\\\label{EApprox}
	E&=4p^2\pi^2\sqrt{K}v(2\pi)\sqrt{1-e^{-4v^2(2\pi)-2I_0}},
\end{align}
where $I_0$ can be replaced with \eqref{I0Approx}.

\section{Thermodynamics}\label{Thermo}
The analytical results obtained in the previous section allow us to define the partition function of the model and study the thermodynamical properties.

As already discussed in Section \ref{Sec:BC}, the baryonic charge $B$ does not take integer values, due to the impossibility for the map $U$ to define a closed cycle. In the following sections, we analyze the case in which $B$ is an integer, by imposing this condition \textit{by hands}, and the solutions when $B$ are continuous.

\subsection{\label{UnitsOfMeasures}A preliminary discussion on the units of measure}

Before introducing the analysis of the thermodynamics, it is useful to fix the order of magnitude of the quantities we are considering. This will be necessary for a numerical estimation of the thermodynamical quantities and a comparison with the real data. 

In this regard, an observation is in order. The model developed in this manuscript can be considered to be at \textit{a preliminary stage} in the following sense. The solutions obtained by our analysis are suitable for describing crystals of baryons organized in layers, which remind of the \textit{lasagna phase} of nuclear matter in neutron stars. Nevertheless, the physics of neutron stars is more complicated than the description given through our model. For instance, nuclear pasta is immersed in a cloud composed by nucleons in its liquid form and a gas of electrons (see for instance \cite{lopez2021properties}). Here, we neglect the interaction with this surrounding matter, which is fundamental in order to give the right estimation of the physical quantities. This work is an attempt to put the basis for a consistent treatment of such models, which are otherwise out of reach by means of analytical tools. Further developments will be considered for future works.

With this in mind, we fix at $L_r\simeq 25\ $fm. Due to  the condition \eqref{p-dependence}, the length of the other edges of the box, called $L$, depends on $p$. For this reason, this quantity cannot be fixed. The system's volume is $V=4\frac{2\pi^3p^2L_r}{2K}$.

Furthermore, we use the following conventional constant values
\begin{align}
    \begin{split}
        f_\pi&=2\sqrt{2}\ \text{fm}^{-1}\simeq 180\ \text{MeV}
        \ (K=\frac{f_\pi^2}{4}=2\ \text{fm}^{-2}).\\
    \end{split}
\end{align} 

\subsection{Integer values of $B$}
\begin{figure}[htb!]\centering
	\includegraphics[width=7.3cm]{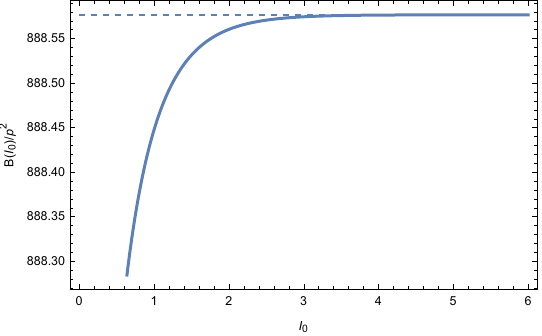}
	\caption{This graph shows the values of $B/p^2$ in terms of $I_0$, where $K=2\ \mbox{fm}^{-2}$, $L_r =25\ \mbox{fm}$. The dashed line represents the maximal value obtained for $I_0\rightarrow\infty$.}
	\label{B/p2}
\end{figure}

We can observe that the quantity $B/p^2$ has a maximal value, depending on the choices of the constant quantities. This constant can be obtained by computing the limit $I_0\rightarrow\infty$ of \eqref{BApprox}. In particular, 
\begin{equation}
	\lim_{I_0\rightarrow\infty}\frac{B}{p^2}=8\sqrt{K}L_r\pi.
\end{equation}
In Fig. \ref{B/p2}, it is represented the behavior of $B/p^2$ and its limit for $I_0\rightarrow\infty$.
\\

As discussed in the previous sections, it is natural to associate an integer value with the baryonic charge. In order to explore this situation, we can consider the values of $v(2\pi)$ for which $B/p^2=n$, where n is an integer; as already mentioned, in our case this has to be imposed by hands. Since $B/p^2$ is bounded, the maximum value of n is $n_{max}=[8\sqrt{K}L_r\pi]$, where the square brackets denote the integer part. For increasing values of $L_r$, $n_{max}$ also increases (in our case $L_r= 25\ \mbox{fm}$, then $n_{max}\approx888.58$). Due to the high number of particles, the analysis of the discrete baryonic charge becomes computationally too expensive. For this reason, in this section consider a smaller value of $L_r$ (i.e., $L_r=1\ \mbox{fm}$, which corresponds to $n_{max}=35$). Subsequently, we extend the study to cases with higher values of $L_r$, adopting a transition to a continuous baryonic charge.

\begin{figure}[htb!]\centering
	\includegraphics[width=7.3cm]{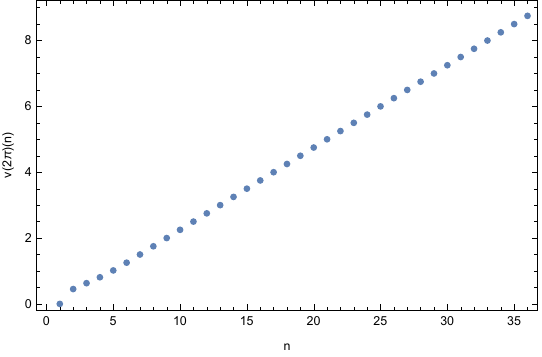}
	\caption{Values of $v(2\pi)(n)$ as a solution of the equation $B/p^2=n$, where $K=2\ \mbox{fm}^{-2}$, $L_r =1\ \mbox{fm}$. For big values of $n$, $v(2\pi)(n)$ enters in a linear regime.}
	\label{I0n}
\end{figure}

The values of $v(2\pi)$ satisfying $B/p^2=n$ when $L_r =1\ \mbox{fm}$ can be determined numerically and are represented in Fig. \ref{I0n}.

Notably, the behavior of $v(2\pi)(n)$ is mostly linear in $n$. This will be useful later.

\subsubsection{The partition function}
For the computation of the grand canonical partition function, 

\begin{equation}
	\mathcal{Z}(\beta, \mu_{B})=\sum_{p,n} e^{-\beta\left(E(p)-\mu_{B} B\right)}
\end{equation}

where $\beta$ is the inverse of the temperature, $E$ is the energy, $\mu_{B}$ is the baryonic chemical potential and $B$ is the baryonic charge.

\begin{figure*}[htb!]\centering

    \begin{subfigure}{0.4\textwidth}
	    \includegraphics[width=\textwidth]{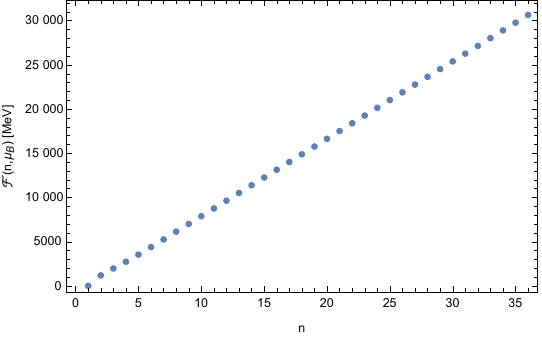}
        \caption{\label{Fig:Fn}}
    \end{subfigure}\hspace{.3cm}
    \begin{subfigure}{0.4\textwidth}
        \includegraphics[width=\textwidth]{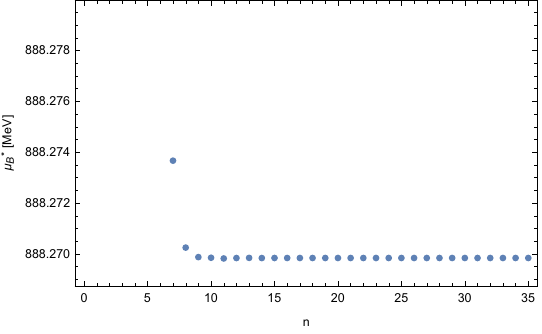}
        \caption{\label{Fig:MuStarn}}
    \end{subfigure}

    \caption{The graphs \ref{Fig:Fn} represents the values of $\mathcal{F}(n,\mu_B)$ for $\mu_{B}=0.2$; notice that, for $n$ big enough, $\mathcal{F}(n,\mu_B)$ is linear in $n$. In the graph \ref{Fig:MuStarn} are reported the values of $\mu_B^*$ as a function of $n$; in this case, $\mu_B^*$ is constant for big values of $n$, as expected. The constants have been fixed to $K=2\ \mbox{fm}^{-2}$, $L_r =1\ \mbox{fm}$.}
	\label{F}
\end{figure*}

\begin{figure*}[htb!]\centering

    \begin{subfigure}{0.4\textwidth}
	    \includegraphics[width=\textwidth]{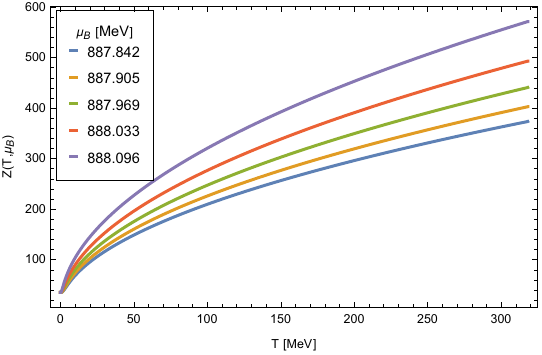}
        \caption{\label{Fig:PT}}
    \end{subfigure}\hspace{.3cm}
    \begin{subfigure}{0.4\textwidth}
        \includegraphics[width=\textwidth]{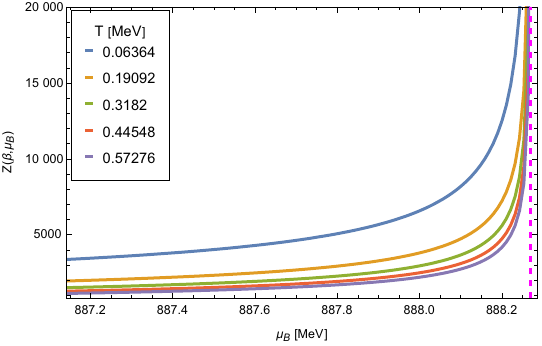}
        \caption{\label{Fig:PMu}}
    \end{subfigure}
    
	\caption{Values of $\mathcal{Z}(T,\mu_B)$ in terms of $T=1/\beta$ and $\mu_B$ , where $K=2\ \mbox{fm}^{-2}$ and $L_r =1\ \mbox{fm}$. The dashed line in the second graph represents the value of $\tilde{\mu}_B$. The dashed magenta line represents the value of $\tilde\mu_B$.}
	\label{Z}
\end{figure*}

In this way, the partition function reads

\begin{align}
	\mathcal{Z}(\beta, \mu_{B})&=\sum_{p,n} e^{-\beta p^2\mathcal{F}(n,\mu_B)},
\end{align}
where $\mathcal{F}(n,\mu_B)$ is the free energy in units of $p^2$. Replacing the energy and the baryonic charge in the exponent of the partition function, we obtain the explicit form of the free energy
\begin{equation}\label{Fn}
	\mathcal{F}(n,\mu_B)=2\pi^2\sqrt{{n}K}\sqrt{v(2\pi)(n)}-n\mu_B,
\end{equation}
represented in the first graph of Fig. \ref{F}.
Notice that $u(2\pi)$ is defined in terms of $n$. The sum over $p$ is convergent only when $\mathcal{F}(n,\mu_B)>0$. This condition is achieved when 
\begin{equation}\label{muStar}
	\mu_{B}<\mu_{B}^*(n),\quad\mbox{with}\quad\mu_B^*(n)=\frac{\mathcal{F}(n,\mu_B=0)}{n}.
\end{equation}
The values of $\mu_{B}^*$ are represented in the second graph of Fig. \ref{F}. Since $\mu_{B}^*(n)$ increases with $n$, in order to satisfy $\mathcal{F}(n,\mu_B)>0$ for each $n$, we should consider the biggest value of $\mu_{B}^*(n)$ (called $\tilde{\mu}_{B}=\mu_{B}^*(n_{max})$) and set $\mu_B<\tilde{\mu}_{B}$.
In this way, the sum leads to the well known theta function named $\theta_3$ (see Appendix \ref{App:Theta} and \cite{whittaker2020course}). In order to simplify the notation, let us introduce the function
\begin{align}\label{xiFunction}
    \xi\left(\tau\right)=\theta_3(e^{i\pi\tau}).
\end{align}
Therefore,
\begin{equation}\label{Theta}
	\sum_{p=-\infty}^{\infty} e^{-\beta p^2\mathcal{F}(n,\mu_B)}=\xi\left(\frac{i\beta}{\pi}\mathcal{F}(n,\mu_B)\right).
\end{equation}
Notice that the quantity \eqref{Theta} diverges in $n=0$. Indeed, $\mathcal{F}(0,\mu_B)=0$. Thus, we define the Zeta function as a sum over $n=1,\dots,n_{max}$. This is equivalent to considering only the states with non-trivial baryonic charge. Then, we have
\begin{equation}\label{ZetaTheta}
	\mathcal{Z}(\beta, \mu_{B})=\sum_{n=1}^{n_{max}}\xi\left(\frac{i\beta}{\pi}\mathcal{F}(n,\mu_B)\right).
\end{equation}
This series clearly has a finite value, since $n$ is limited. The behavior of $\mathcal{Z}(\beta,\mu_B)$ in terms of $\beta$ and $\mu_B$, for fixed values of $\mu_B$ and $\beta$ respectively, is represented in Fig. \ref{Z}.

It is worth making the following observations. We remember that when $n$ is big, thus approaching $n_{\text{max}}$, the quantity $I_0\rightarrow\infty$. Therefore, 
\begin{equation}
	1-\exp\left(-8\sqrt{K}L_r\pi+4v(2\pi)\right)\rightarrow 0
\end{equation}
and, from \eqref{BApprox}, we observe that $v\propto n$. In particular,
\begin{equation}
	v(2\pi)\rightarrow\frac{n}{4}.
\end{equation}
It becomes evident that in this regime also $E\propto n$. This behavior can be observed in the graph \ref{I0n}, suggesting that also $\mathcal{F}(n,\mu_B)\propto n$ for $n$ large enough (see Fig. \ref{F}). In particular, in this limit $\mathcal{F}(n,\mu_B)$ can be approximate as
\begin{equation}
	\mathcal{F}(n,\mu_B)\approx n\left[\tilde\mu_{B}-\mu_{B}\right],
\end{equation}
with $\tilde\mu_B=\pi^2K$. When $\mu_B\approx\tilde\mu_B$, then $\mathcal{F}(n,\mu_B)\approx 0$ and the theta function diverges. The degree of the divergence can be studied through the identity (see Appendix \ref{App:Theta})
\begin{equation}
    \xi(\tau)=\frac{1}{\sqrt{-i\tau}}\xi\left(- \frac{1}{\tau} \right).
\end{equation}
In our case, $\tau=\frac{i\beta}{\pi}\mathcal{F}(n,\mu_B)$. Therefore, the function $\xi(\tau)$ can be written as
\begin{align}\hspace{-2.5pt}
    \xi\left(\frac{i\beta}{\pi}\mathcal{F}(n,\mu_B)\right)
    =\frac{1}{\sqrt{\frac{\beta}{\pi}\mathcal{F}(n,\mu_B)}}\left(1+2\sum_{p=1}^{\infty} q^{p^2}\right).
\end{align}
where $q=\exp{\left(-\frac{\pi2}{\beta\mathcal{F}(n,\mu_B)}\right)}$. When $\beta(\tilde\mu_{B}-\mu_{B})$ is small enough, then $q\approx 0$ and 
\begin{equation}\label{ThetaApprox}
    \xi\left(\frac{i\beta}{\pi}\mathcal{F}(n,\mu_B)\right)\approx\frac{1}{\sqrt{\frac{\beta}{\pi}n\left[\tilde\mu_{B}-\mu_{B}\right]}}.
\end{equation}

Notably, this approximation is valid also when $n$ is small. In this case, $\tilde{\mu}_B$ is replaced by $\mu_B^*(n)$ and $\beta$ is small enough.

This divergence indicates that $\widetilde{\mu }_{B}$ is the maximal value of the chemical potential, beyond which the present computations cannot be trusted (see also the analysis in section \ref{Thermo} for a further discussion). It is worth emphasizing that obtaining a finite range for the baryonic chemical potential through an explicit analytic expression is a significant result, particularly given the well-known challenges Lattice QCD faces at finite baryon densities.

\subsection{\label{Sec:ContinuousB}Continuous values of $B$}
As already mentioned, considering the high number of particles, the assumption that $B$ takes only integer values can be overcome. Indeed, our model allows $n$ to vary continuously from $0$ to $n_{max}$. This could be interpreted as due to a quantum screening of the baryonic charge so making it a continuous number. Furthermore, the analysis with a continuous baryonic charge becomes particularly relevant when $L_r$ becomes large. For instance, the case analyzed in this manuscript considers $L_r=25$, corresponding to $n_{max}\approx888.58$. The properties of the partition function analyzed in the previous section can be adapted to the continuous case, with the following observations.

\begin{figure*}[htb!]\centering

    \begin{subfigure}{0.4\textwidth}
	    \includegraphics[width=\textwidth]{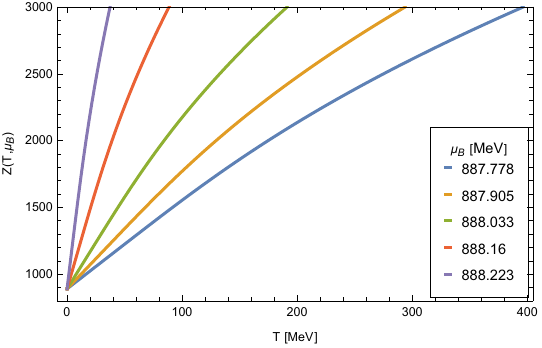}
        \caption{\label{Fig:CPT}}
    \end{subfigure}\hspace{.3cm}
    \begin{subfigure}{0.4\textwidth}
        \includegraphics[width=\textwidth]{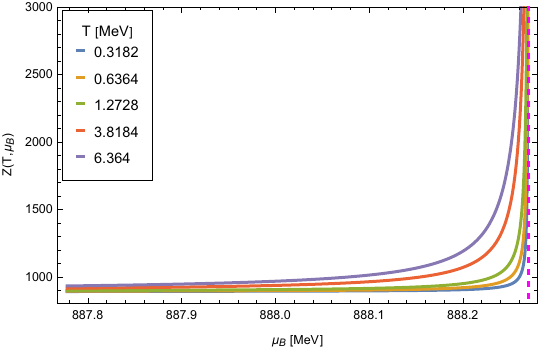}
        \caption{\label{Fig:CPMu}}
    \end{subfigure}
    
	\caption{Values of $\mathcal{Z}(T,\mu_B)$ in terms of $T=1/\beta$ and $\mu_B$ , where $K=2\ \mbox{fm}^{-2}$ and $L_r =25\ \mbox{fm}$. The dashed line in the second graph represents the value of $\tilde{\mu}_B$.}
	\label{CZ}
\end{figure*}

When $n$ is continuous, the partition function reads
\begin{align}
	\mathcal{Z}(\beta, \mu_{B})=\int_0^{n_{max}}\xi\left(\frac{i\beta}{\pi}\mathcal{F}(n,\mu_B)\right)\ dn,
\end{align}
where now $n_{max}=8\sqrt{K}L_r\pi$ is not an integer. Its behavior in terms of the temperature $T$ and $\mu_B$ is represented in Fig. \ref{CZ}
Notice that the theta function diverges as $n^{-\frac 12}$ in $n=0$, so the integral converges.

Interestingly, this integral leads to a nice connection with the Riemann zeta function when $n_{max}$ is large. Indeed, if we introduce the function 
\begin{align}
    \Theta(x)=\xi(ix),
\end{align}
it holds
\begin{align}
    \int_0^\infty x^{\frac s2-1} (\Theta(x)-1)dx=2 \frac {\Gamma(s/2)}{\pi^{s/2}} \zeta(s),
\end{align}
for any $s$ such that $Re(s)>0$. In this situation, assuming $\beta(\tilde\mu_{B}-\mu_{B})$ not too small, we can approximate ${\mathcal F(n,\mu_B)\approx n\left[\tilde\mu_{B}-\mu_{B}\right]}$. Furthermore, let us define  
\begin{align}
    \tilde{\mathcal{Z}}(\beta, \mu_{B})=\mathcal{Z}(\beta, \mu_{B})-n_{max}.
\end{align}
Thus, 
\begin{align}\label{ZApproxBig}
    \tilde{\mathcal{Z}}(\beta, \mu_{B})&\simeq \int_0^\infty\Big[\Theta\left(\frac{\beta}{\pi}\mathcal{F}(n,\mu_B)\right)-1\Big]\ dn\cr
    &=\frac 2\beta \frac {\zeta(2)}{\tilde\mu_{B}-\mu_{B}}=\frac {\pi^2}{3\beta(\tilde\mu_{B}-\mu_{B})}.
\end{align}
More generally, one can compute the expectation value of an extensive quantity $Q(n)=w n^s$ as
\begin{align}
    \langle Q(n) \rangle&\simeq \frac w{\mathcal{Z}(\beta, \mu_{B})} \int_0^\infty\Big[\Theta_3\left(\frac{\beta}{\pi}\mathcal{F}(n,\mu_B)\right)-1\Big]n^s\ dn\cr
    &=\frac {w}{\beta^s}\frac 1{(\tilde\mu_{B}-\mu_{B})^s} {\Gamma(s+1)} \frac {\zeta(2s+2)}{\zeta(2)}.
\end{align}
This formula has two kinds of corrections. One is due to the finiteness of $n_{\max}$ that gives corrections of order $e^{-\beta n_{max}(\tilde\mu_{B}-\mu_{B})}$. 
It cannot be used for small $\beta n_{max}(\tilde\mu_{B}-\mu_{B})$. The other corrections come from the fact that the behavior of $\mu^*_B$ changes for small $n$.
Indeed, using equations \eqref{Fn} and \eqref{muStar},  we can explicitly write
\begin{gather}
    \mathcal{F}(n,\mu_B)= n\left[\mu^*_{B}(n)-\mu_{B}\right],\\
    \mu^*_{B}(n)=2\pi^2  K^{\frac{1}{2}} \sqrt{\frac{v(2\pi)(n)}{n}}.
\end{gather}

In order to find $v(2\pi)$ in terms of $n$, we need to solve the equation $B/p^2=n$. In this scope, let us consider equation \eqref{BApprox} with the approximations \eqref{I0Approx}. The explicit expression for $B/p^2$ is

\begin{widetext}
\begin{equation}
    \frac{B}{p^2}\approx\left\{\begin{array}{l r}
	      4v(2\pi)\left[1-e^{-4v^2(2\pi)}(1-e^{4v(2\pi)-8\pi\sqrt{K}L_r})\right]\qquad & \mbox{for }v(2\pi)\geq\tilde\varepsilon,\\
	       4v(2\pi)\left[1-e^{-4v^2(2\pi)\left(1+\frac{1}{\sinh^2{(4\pi\sqrt{K}L_r)}}\right)}\right]\qquad & \mbox{for }v(2\pi)<\tilde\varepsilon.
	\end{array}\right.
 \end{equation}
\end{widetext}

\begin{figure*}[htb!]\centering

    \begin{subfigure}{0.4\textwidth}
	    \includegraphics[width=\textwidth]{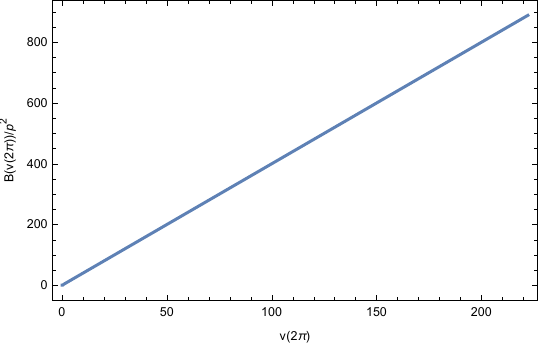}
        \caption{\label{Fig:BVSv}}
    \end{subfigure}\hspace{.3cm}
    \begin{subfigure}{0.4\textwidth}
        \includegraphics[width=\textwidth]{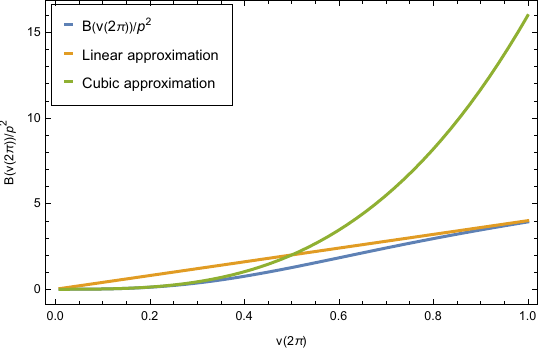}
        \caption{\label{Fig:BVSvZoom}}
    \end{subfigure}
    
	\caption{Values of $B/p^2=n$ in terms of $v(2\pi)$, where $K=2\ \mbox{fm}^{-2}$ and $L_r =25\ \mbox{fm}$. The graph \ref{Fig:BVSv} represents the relation in the whole range of $v(2\pi)$. The graph \ref{Fig:BVSvZoom} represent the same relation for small $(2\pi)$, compared to the approximation \eqref{vsmallApprox} and the linear approximation $n=v(2\pi)/4$.}
	\label{Linearv}
\end{figure*}

Since the second relation is valid for very small values of $v(2\pi)$, in this limit we can write
\begin{equation}
    \frac{B}{p^2}\approx 16v^3(2\pi)\left(1+\frac{1}{\sinh^2{(4\pi\sqrt{K}L_r)}}\right).
\end{equation}
Then,
\begin{equation}\label{vsmallApprox}
    v(2\pi)\approx\left[\frac{n}{16\left(1+\sinh^{-2}{(4\pi\sqrt{K}L_r)}\right)}\right]^{\frac{1}{3}}
\end{equation}
for very small $v(2\pi)$. The precise result for $v(2\pi)$ not so small needs to be found numerically. Nevertheless, the picture \eqref{Linearv} shows that the behavior of $n$ in terms of $v(2\pi)$ (and, thus, the vice versa) is mostly linear when $n_{max}$ is big. In particular, from the graph \ref{Fig:BVSvZoom} it is clear that the linear approximation is valid from small $v(2\pi)$ (around $0.8$, corresponding to $n\approx 3.2$).

In order to simplify the computations, in what follows we use the linear approximation for $v(2\pi)$.

\subsubsection{Thermodynamical quantities}
The explicit form of the partition function allows us to compute several thermodynamical quantities in terms of the temperature $T$ and $\mu_B$. The quantities analyzed in this section are the following

\textbf{Internal energy:}
\begin{equation}
    U = -\frac{\partial \ln{\mathcal{Z}}}{\partial\beta};
\end{equation}
\textbf{Entropy:}
\begin{equation}
    S = k_B\left[\ln\mathcal{Z}-\beta\frac{\partial \ln{\mathcal{Z}}}{\partial\beta}\right];
\end{equation}
\textbf{Average number of particles:}
\begin{equation}
    \langle N\rangle = \frac{1}{\beta}\frac{\partial \ln{\mathcal{Z}}}{\partial\mu_B};
\end{equation}
\textbf{Heat capacity:}
\begin{equation}
    C_V = k_B\beta^2\ \frac{\partial^2 \ln{\mathcal{Z}}}{\partial\beta^2}.
\end{equation}

\begin{figure*}[htb!]\centering

    \vspace{0.5cm}
    \textbf{Average number of particles}
    \vspace{0.5cm}

    \begin{subfigure}{0.4\textwidth}
        \includegraphics[width=\linewidth]{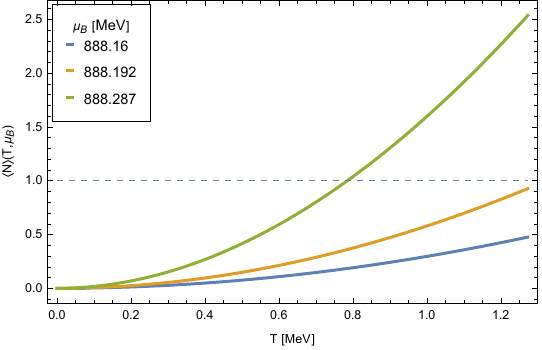}
        \caption{\label{Fig:NPT}}
    \end{subfigure}\hspace{.3cm}
    \begin{subfigure}{0.4\textwidth}
        \includegraphics[width=\linewidth]{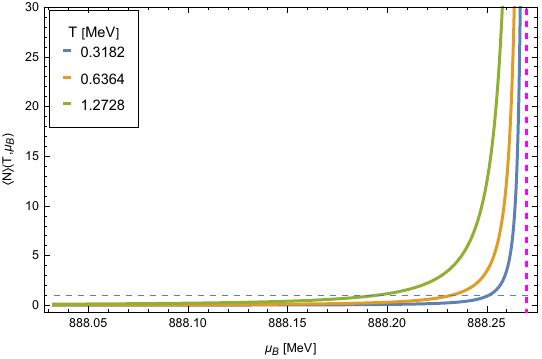}
        \caption{\label{Fig:NPMuB}}
    \end{subfigure}

    \caption{Average numbers of particles as a function of the temperature $T$ (Fig. \ref{Fig:NPT}) and baryonic chemical potential $\mu_B$ (Fig. \ref{Fig:NPMuB}), with $K=2\ \mbox{fm}^{-2}$ and $L_r =25\ \mbox{fm}$. The dashed horizontal line represents the case with $\langle N\rangle=1$.}
    \label{fig:NP}
\end{figure*}

\begin{figure*}[htb!]\centering

    \vspace{0.5cm}
    \textbf{Internal energy}
    \vspace{0.5cm}

    \begin{subfigure}{0.4\textwidth}
        \includegraphics[width=\linewidth]{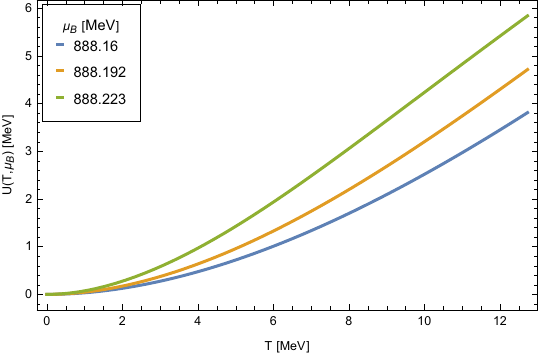}
        \caption{\label{Fig:IET}}
    \end{subfigure}\hspace{.3cm}
    \begin{subfigure}{0.4\textwidth}
        \includegraphics[width=\linewidth]{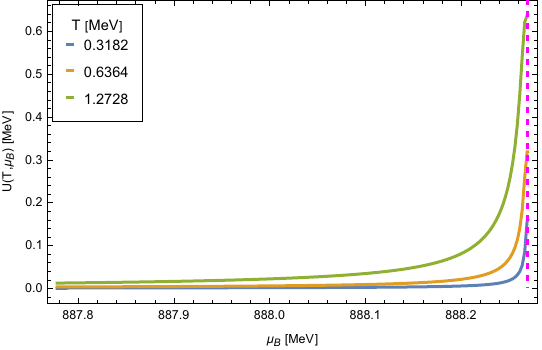}
        \caption{\label{Fig:IEMuB}}
    \end{subfigure}

\caption{Internal energy as a function of the temperature $T$ (Fig. \ref{Fig:IET}) and baryonic chemical potential $\mu_B$ (Fig. \ref{Fig:IEMuB}), with$K=2\ \mbox{fm}^{-2}$ and $L_r =25\ \mbox{fm}$.}
\label{fig:IE}
\end{figure*}

\begin{figure*}[htb!]\centering

    \vspace{0.5cm}
    \textbf{Entropy}
    \vspace{0.5cm}

    \begin{subfigure}{0.4\textwidth}
        \includegraphics[width=\linewidth]{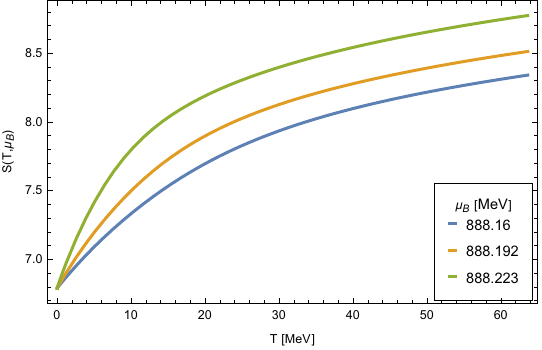}
        \caption{\label{Fig:ST}}
    \end{subfigure}\hspace{.3cm}
    \begin{subfigure}{0.4\textwidth}
        \includegraphics[width=\linewidth]{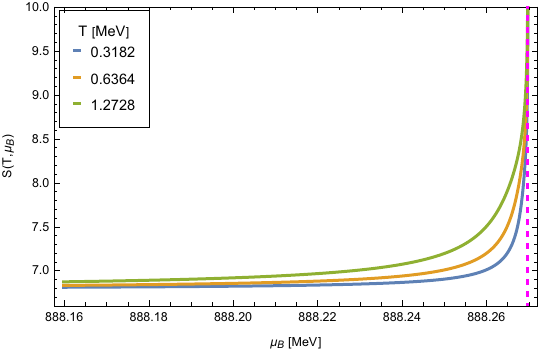}
        \caption{\label{Fig:SMuB}}
    \end{subfigure}

    \caption{Entropy as a function of the temperature $T$ (Fig. \ref{Fig:ST}) and baryonic chemical potential $\mu_B$ (Fig. \ref{Fig:SMuB}), with $K=2\ \mbox{fm}^{-2}$ and $L_r =25\ \mbox{fm}$.}
    \label{fig:S}
\end{figure*}

\begin{figure*}[htb!]\centering

    \vspace{0.5cm}
    \textbf{Heat capacity}
    \vspace{0.5cm}

    \begin{subfigure}{0.4\textwidth}
        \includegraphics[width=\linewidth]{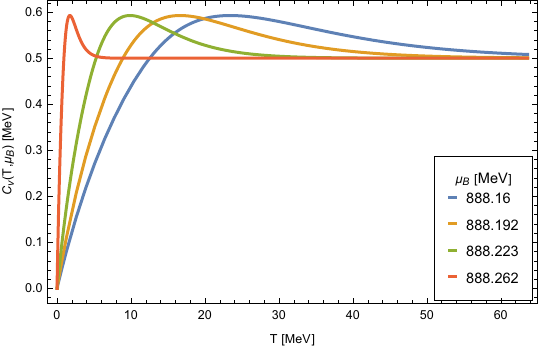}
        \caption{\label{Fig:HCT}}
    \end{subfigure}\hspace{.3cm}
    \begin{subfigure}{0.4\textwidth}
        \includegraphics[width=\linewidth]{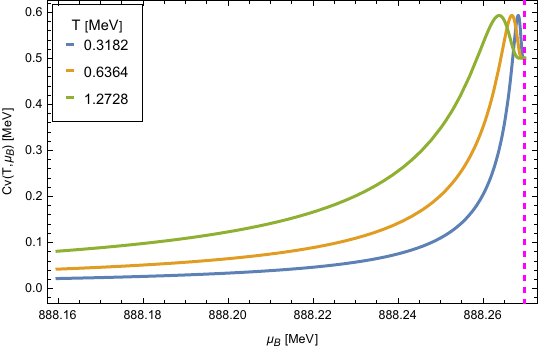}
        \caption{\label{Fig:HCMuB}}
    \end{subfigure}
    
	\caption{Heat capacity as a function of the temperature $T$ (Fig. \ref{Fig:HCT}) and baryonic chemical potential $\mu_B$ (Fig. \ref{Fig:HCMuB}), with $K=2\ \mbox{fm}^{-2}$ and $L_r =25\ \mbox{fm}$.}
    \label{fig:HC}
\end{figure*}

Remember that $k_B$ is the Boltzmann constant and $\beta=1/(k_B T)$, where $T$ is the temperature. Here, we consider $k_B=1$. The relative plots are reported in Figs. \ref{fig:IE}, \ref{fig:S}, \ref{fig:NP} and \ref{fig:HC}. In some of them, very high temperatures have been considered in order to show the behavior of the plot.

The plots in the picture \ref{fig:NP} are particularly interesting, since they highlight a lower bound for the temperature $T$ at different values of $\mu_B$ (plot \ref{Fig:NPT}) or a lower bound for $\mu_B$ at different values of $T$ (plot \ref{Fig:NPMuB}). Indeed, looking at the plot \ref{Fig:NPT}, once the baryonic chemical potential is fixed, the temperature must exceeds a threshold ($T\geq T_0(\mu_B)$) in order to produce at least one particle. Vice versa, when $T$ is fixed, then $\mu_B$ should be higher then a certain value ($\mu_B\geq \mu_{B,0}(T)$). 

Notably, for $\tilde\mu_B-\mu_B$ sufficiently small, the range of $T$ overlaps the one associated to the nuclear pasta formation, which is typically $0.5\ \mbox{MeV}\leq T \leq 15\ \mbox{MeV}$ \cite{lopez2021properties,Xia:NPSHighTemp}. This behavior can be directly observed from the approximation \eqref{ThetaApprox}. Indeed, when $\tilde\mu_B-\mu_B$ is very small, then
\begin{align}
    \langle N\rangle\approx\frac{1}{2\beta(\tilde\mu_B-\mu_B)}.
\end{align}
Thus, $\langle N\rangle> 1$ as $T> 2(\tilde\mu_B-\mu_B)$, considering $k_B=1$. The lower bound of $T$ becomes zero as $\mu_B$ approaches $\tilde\mu_B$. Nonetheless, we need to take into account that $T$ cannot be equal to its lower bound, otherwise the approximation \eqref{ThetaApprox} is no longer valid. Furthermore, the case with $\mu_B=\tilde\mu_B$ causes some thermodynamical quantities to diverge. In  conclusion, it is not possible to study the behavior of the system at $T=0$ MeV, but it is possible to set $\mu_B$ in order to reach very low temperatures, in a range associated to nuclear pasta formation. Furthermore, we need to point out that the values of $\tilde\mu_B\approx888.27$ MeV, which is lower that the baryonic chemical potential found for nuclear pasta formation, but not so far. Indeed, it typically in the range of $930-1000$ MeV (see, for instance, \cite{Schmitt:BCP}). 

As already observed in the case of integer baryonic charge, the limit of $\tilde\mu_B$ cannot be passed, and the divergent behavior of some thermodynamical can be associated to a phase transition, but we need to take into account that our model is suitable for describing only one particular state of nuclear matter (i.e., layers of baryons) and the possibility of other phases cannot be investigated. Thus, $\tilde\mu_B$ has to be considered as a limit of validity for the model in question.

Noteworthy, this model does not provide an upper bound for the temperature, which is generally associated to phase transition. Therefore, the structures studied in this paper exist also for very high temperatures.

Observing the plots in Fig. \eqref{fig:S}, one may notice that the entropy does not go to zero for $T=0$. This is due to the presence of a degenerate state for $p=0$. Indeed, when $T$ is very small (and thus $\beta\rightarrow\infty$), the only surviving term in the partition function is represented by the fundamental state at $p=0$. This state is also characterized by a null volume, due to the relation \eqref{p-dependence} imposed in order to obtain the BPS conditions, and zero energy. Nonetheless, it contains $0\leq n\leq n_{max}$ layers, leading to a degeneracy. In particular,
\begin{align}
    \lim_{\beta\rightarrow\infty}\mathcal{Z}=\int_{n=0}^{n_{max}}1\ dn=n_{max}.
\end{align}
Then, the entropy reads
\begin{align}
    \lim_{\beta\rightarrow\infty}S=k_B\ln(n_{max}).
\end{align}

Finally, the heat capacity $C_v$, shown in Fig. \ref{fig:HC}, exhibits a characteristic \textit{Schottky-like anomaly} \cite{kittel2018introduction}. The position of the peak can be computed analytically through the approximations \eqref{ThetaApprox} for $\beta(\tilde\mu_B-\mu_B)\rightarrow0$ and \eqref{ZApproxBig} for finite values of $\beta(\tilde\mu_B-\mu_B)$. Interestingly, for certain values of $\mu_B$, the peak temperatures occur around $T\approx 10-15$ MeV, a range generally associated with the phase transition from pasta-like structures to uniform nuclear matter \cite{Caplan:ThermalFluctuation}. As $\mu_B\rightarrow\tilde\mu_B$, this peak shifts toward $T\rightarrow0$. Although the absence of a singularity (typically appearing as a $\lambda$-peak) suggests the lack of a formal phase transition, this behavior warrants further investigation in future work.


\section{External field and magnetic susceptibility}\label{MagSusc}

\begin{figure*}
    \centering\hspace{.3cm}
    \includegraphics[width=0.35\textwidth]{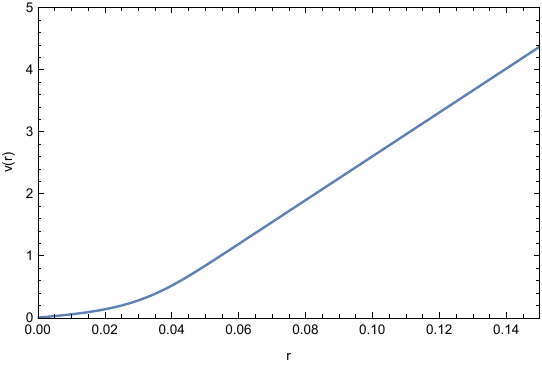}\hspace{.3cm}
    \includegraphics[width=0.37\textwidth]{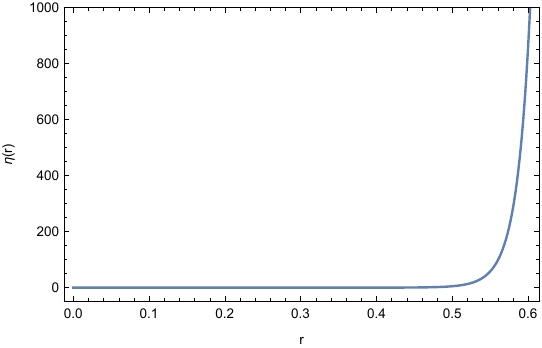}
    \caption{Numerical solution for $v(r)$ and $\eta(r)$, where $K=2\ \mbox{fm}^{-2}$, $L_r=L =25\ \mbox{fm}$. and $I_0=0.01$}
    \label{fig:u(r)}
\end{figure*}

\begin{figure*}
    \centering
    \includegraphics[width=0.36\textwidth]{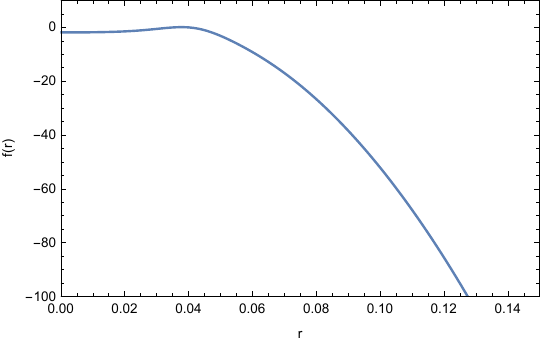}\hspace{.3cm}
    \includegraphics[width=0.36\textwidth]{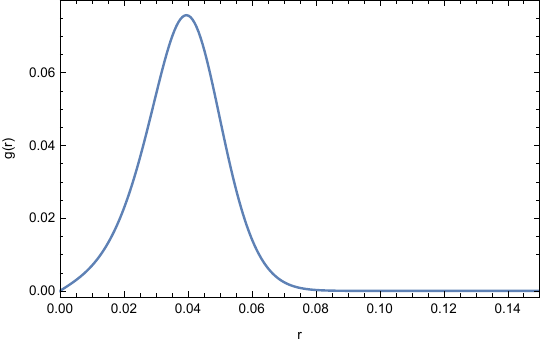}
    \caption{Numerical solution for $f(r)$ and $g(r)$, where $K=2\ \mbox{fm}^{-2}$, $L_r=L =25\ \mbox{fm}$. and $I_0=0.01$}
    \label{fig:f(r)}
\end{figure*}

Let us now discuss the possibility of introducing an external small perturbation of the magnetic field. In particular,
\begin{equation}\label{Pert1}
	v(r)\rightarrow v(r)+b_e r,
\end{equation}
where $b_e$ is arbitrarily small and represents the magnitude of the external magnetic field. Thus, also $H(r)$ varies by a small quantity. Hence
\begin{equation}\label{Pert2}
	H(r)\rightarrow H(r)+I(r;b_e),
\end{equation}
where $I(r;b_e)$ is a $b_e$-dependent function of r, which has to be determined. It undergoes the condition 
\begin{equation}\label{CondOnI}
	I(r;0)=0.
\end{equation}
Since $b_e$ is small, we can consider $I(r;b_e)$ computed at the first order in $b_e$. Namely,
\begin{equation}
	I(r;b_e)\simeq b_e I'(r;0)=b_e\eta(r),
\end{equation}
where the \textit{prime} indicated the derivative in $b_e$. Here, we used the condition \eqref{CondOnI}. This way, the equation of motion for $H$ \eqref{eq1}, at the first order in $B$, becomes
\begin{widetext}
\begin{align}\label{eqPert1}
	H''+4p^2\left(\frac{L_r}{L}\right)^2 \sin(2H)\left(\frac{1}{4}-v^2\right)+b_e\left\{\eta''+8\left(\frac{L_r}{\bar L}\right)^2\left[\eta\cos(2H)\left(\frac{1}{4}-v^2\right)-vr\sin(2H)\right]\right\}=0,
\end{align}
\end{widetext}
where, now, the \textit{prime} represents the derivative in $r$, and we introduced $\bar L=L/p$. The functions $H$ and $u$ are solutions of the BPS equations \eqref{Eq1} and \eqref{Eq2}. Thus, the part of the equation \eqref{eqPert1} which is not proportional to $B$ is automatically solved. We are now left with an equation for the perturbation $\eta$. 
We can use the relations \eqref{H} in order to express $H$ in terms of $v$. Namely,
\begin{align}
    \cos(2H)&=1-2 e^{-4v^2-2I_0 } \\
    \sin(2H)&=2e^{-2v^2+2I_0  }\sqrt{1-e^{-4v^2-2I_0 }}
\end{align}
The equation for $\eta$ can be written as
\begin{equation}\label{eta_ODE}
    \eta''+f\eta=g,
\end{equation}
where
\begin{align}
    \hspace{-6pt}f(v,r)&=2\left(\frac{L_r}{\bar L}\right)^2 \left ( 1-2 e^{-4v^2-2I_0 }\right)\left(1-4v^2\right),\\
    \hspace{-6pt}g(v,r)&=8\left(\frac{L_r}{\bar L}\right)^2 v r\left(2e^{-2 v^2+2I_0  }\sqrt{1-e^{-4 v^2-2I_0 }} \right).
\end{align}
We need to remark that in order to find a solution for $\eta$, it is also necessary to numerically solve $v(r)$ and, consequently, $f(r)$ and $g(r)$. The numerical solutions have been reported in Figs. \ref{fig:u(r)} and \ref{fig:f(r)}. Notice that for $I_0 \leq 1$, $v(r) \approx r$, and as $I_0$ becomes smaller, it enters a linear regime. This agrees with the fact that for small $I_0$, we enter the low-energy regime and, therefore, the highly interactive regime.

\vspace{1cm}
\subsection{The Maxwell equations}
The perturbation of the external field $B$ also affects the Maxwell equation \eqref{eq2}, which represents the equation of motion for the magnetic field $u(r)$. This case requires the generation of an external current $J_e$, which contributes as follows
\begin{equation}\label{eq2Pert}
	\hat{v}''-4KL_{r}^{2}\sin^2(\hat{H})\hat{v}=J_e ,
\end{equation}
where $\hat{v}(r)=v(r)+b_er$ and $\hat{H}(r)=H(r)+b_e\eta$ are the perturbed solutions. At first order approximation, \eqref{eq2Pert} becomes
\begin{align}
    J_e&=
	\left[v''-4KL_{r}^{2}\sin^2(H)v\right]\cr
    &\qquad\qquad-4b_eKL_r^2\left(r\sin^2H+\eta v\sin(2H)\right).
\end{align}
The part in square brackets is solved by the usual solution of equation \eqref{eq2}. This is allowed if the external current is
\begin{equation}
	J_e=-4b_eKL_r^2\left(r\sin^2H+\eta v\sin(2H)\right).
\end{equation}

\subsection{Contribution to the energy density and total energy}
\begin{figure}[h!]
	\centering
	\includegraphics[width=7.3cm]{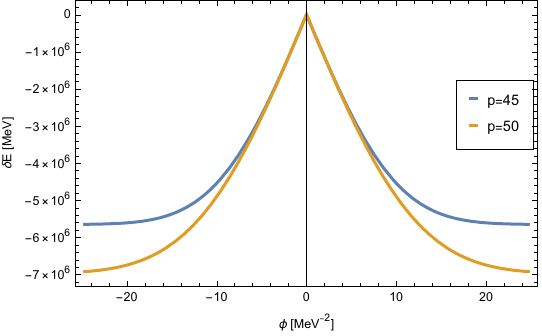}
	\caption{Contribution of the external field to the total energy in terms of the flux $\Phi$ where $K=2\ \mbox{fm}^{-2}$ and $L_r=25\ \mbox{fm}$.}
	\label{fig:dE}
\end{figure}

A straightforward computation leads to the first-order contribution of the external field to the energy density \eqref{EnDensity}
\begin{widetext}
\begin{equation}
	T_{00}^{b_e}=T_{00}+b_e\left[\frac{8K}{\bar L^2}vr\sin^2H+\frac{4K}{\bar L^2}\eta\left(v^2-\frac{1}{4}\right)\sin(2H)+\frac{K}{ L_r^2}\eta'H'+\frac{2v'}{(L_r \bar L)^2}\right].
\end{equation}
\end{widetext}
Using equations \eqref{eq1} and \eqref{eq2}, $T_{00}^{b_e}$ can be rewritten as
\begin{equation}
	T_{00}^{b_e}=T_{00}+\frac{b_e}{L_r^2}\frac{d}{dr}\left(\frac{2v'r}{\bar L^2}+K\eta H'\right).
\end{equation}
The contribution to the energy density is thus represented by a total derivative. The quantities $u'$ and $H'$ can be found using the BPS equations \eqref{Eq1} and \eqref{Eq2}, {remembering the condition \eqref{p-dependence}}. More explicitly,
\begin{equation}\label{EnDensityPert}
 	T_{00}^{b_e}=T_{00}+\frac{4K^{\frac{3}{2}}b_e}{L_r}\frac{d}{dr}\left({r}\cos H-v\eta \sin H\right).
\end{equation}
This result allows us to compute the contribution to the total energy, obtained by integrating \eqref{EnDensityPert} over the volume
\begin{align}
	E_{b_e}=E+b_e\delta E,
\end{align}
where
\begin{align}\label{DeltaE}
    \delta E&={8\pi^3K^{\frac{1}{2}}}\cos H(2\pi)\cr
    &\qquad\quad-{9\pi^2p^2K^{\frac{1}{2}}}v(2\pi)\eta(2\pi)\sin H(2\pi).
\end{align}
The contribution of the external field to the energy can be rewritten in terms of the flux $\Phi$ and the constant $I_0$ using equations \eqref{H} and \eqref{Phiu}. Then, a relation between $\Phi$ and $I_0$ can be obtained as usual by solving the integral \eqref{equation} numerically or considering the approximation \eqref{ApproxSol} for the analytical results. In order to understand the behavior of the energy when an external magnetic field is present, we need to define the boundary condition $\eta(2\pi)$. In this scope, we should compute the contribution of the external field to the baryonic charge. 
\begin{equation}
	B_{b_e}=B+b_e\delta B,
\end{equation}
where
\begin{align}
    \delta B&=-12p^2b_e\left[4\pi(1-\sin^2(H(2\pi)))\right.\notag\\
    &\qquad\left.-4\eta(2\pi)v(2\pi)\sin(H(2\pi))\cos(H(2\pi))\right].
\end{align}
The effects of the external field should not spoil the value of the baryonic charge. Thus,
\begin{equation}
	4\pi(1-\sin^2(H(2\pi)))-4\eta(2\pi)v(2\pi)\cos(H(2\pi))=0,
\end{equation}
which solution is
\begin{align}
    \eta(2\pi)=\frac{\pi}{v(2\pi)}\frac{\cos(H(2\pi))}{\sin(H(2\pi))}
\end{align}
Then, the contribution \eqref{DeltaE} brought by the external field to the total energy becomes
\begin{equation}
    \delta E = -{\pi^3p^2K^{\frac{1}{2}}}\cos (H(2\pi)).
\end{equation}
Fig. \ref{fig:dE} represents the contribution of the external field to the total energy computed through the approximated solution \eqref{ApproxSol}.
{It is worth noticing here that when $b_{e}>0$ (namely, when the external field is parallel to the magnetic field of the layer itself), then  $\delta E<0$ and the solutions analyzed in this paper present a ferromagnetic behavior.}

The final contribution at the first order in $b_e$ to the free energy can be written as
\begin{equation}\label{Fbe}
    \mathcal{F}_{b_e}=\mathcal{F}+b_enp^2\ \delta\mu_B,
\end{equation}
where
\begin{equation}
    \delta\mu_B = \frac{1}{np^2}\left(\delta E- L_r\frac{\partial\delta E}{\partial L_r}\right).
\end{equation}

\subsection{Magnetic susceptibility}
\begin{figure}[h!]
    \centering
    \includegraphics[width=7.3cm]{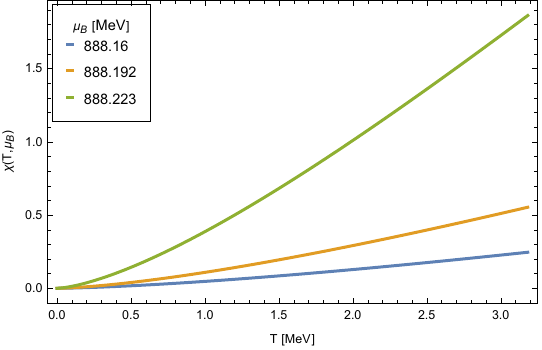}
    \caption{Magnetic susceptibility in terms of the temperature $T$ for different values of $\mu_B$, where $K=2\ \mbox{fm}^{-2}$, $L_r =25\ \mbox{fm}$.}
    \label{fig:Chi}
\end{figure}

The magnetic susceptibility can be computed analytically from the partition function through
\begin{equation}
    \chi = \frac{\mu_0}{\beta}\underset{b_e\rightarrow 0}{\lim}\left(\frac{\partial^2\ln\mathcal{Z}_{b_e}}{{\partial b_e}^2}\right),
\end{equation}
where $\mu_0$ is the vacuum permeability and $\mathcal{Z}_{b_e}$ is the partition function computed using $\mathcal{F}_{b_e}$ in Eq. \eqref{Fbe}. Its dependence on different values of the temperature $T$ is depicted in the graphs of Fig. \ref{fig:Chi}.

The behavior at small and big temperatures can also be studied analytically by writing the expression of $\chi$ more explicitly as
\begin{align}\label{chiExpand}
    \chi &= \frac{\mu_0\beta}{\pi^2\mathcal{Z}^2}\sum_{n=1}^{n_{max}}\Big[\Delta^2(n)\Theta''(x)|_{x=\frac{\beta}{\pi}x_0(n)}\mathcal{Z}\\\nonumber
    &\quad-\Delta(n)\Theta'(x)|_{x=\frac{\beta}{\pi}x_0(n)}\sum_{m=1}^{n_{max}}\Delta(m)\Theta'(x)|_{x=\frac{\beta}{\pi}x_0(m)}\Big],
\end{align}
where $x_0(n)=n(\mu^*_B(n)-\mu_B)$, $\Delta(n)=n\ \delta\mu_B(n)$ and the \textit{prime} indicates the derivative with respect to $x$. This expression is also valid for continuous values of $n$ by replacing the sum with the integral.

\textit{For big values of $T$} (and, thus, small $\beta$), we can use the approximation \eqref{ApproxTheta} for $\Theta(x)$ in order to get
\begin{align}
    \chi\approx\frac{\mu_0}{4\beta}\sum_{n,m=1}^{n_{max}}\left[\frac{3\Delta^2(n)}{[x_0(n)]^{\frac{5}{2}}[x_0(m)]^{\frac{1}{2}}}-\frac{\Delta(n)\Delta(m)}{[x_0(n)]^{\frac{3}{2}}[x_0(m)]^{\frac{3}{2}}}\right]
\end{align}
Since $\beta=1/(k_BT)$, then, $\chi\propto T$ in this limit.

\textit{When $\beta$ is large} (and $T$ is small), we can use the explicit expression for $\Theta(x)$. Namely,
\begin{equation}
    \Theta(x)=1+2\sum_{p=1}^{\infty}e^{-p^2\pi x}.
\end{equation}
Thus,
\begin{equation}
    \frac{d}{dx}\Theta(x)=-2\pi\sum_{p=1}^{\infty}p^2e^{-p^2\pi x},
\end{equation}
and
\begin{equation}
    \frac{d^2}{dx^2}\Theta(x)=2\pi^2\sum_{p=1}^{\infty}p^4e^{-p^2\pi x},
\end{equation}
where $x=\frac{\beta}{\pi}n(\mu^*_B(n)-\mu_B)$. Then, when $\beta\rightarrow\infty$, both the first and second derivative goes to zero with order $e^{-\beta}$.  Moreover, $\Theta(x)\rightarrow 1$ and $\mathcal{Z}$ is a finite constant different from zero. This means that $\chi\rightarrow 0$ as $\beta\rightarrow\infty$.

Once again, it can be observed that this is also valid when $n$ varies continuously.

The fact that the magnetic susceptibility in Fig. \ref{fig:Chi} grows with temperature can be interpreted as follows. At low temperature, the growing behavior can be associated with the \textit{Hopkinson effect}, which is typical of the ferromagnetic and ferrimagnetic materials \cite{hopkinson1889xiv}. Once the temperature reaches the Curie temperature $T_c$, the magnetic susceptibility should start decreasing. On the other hand, at high temperatures, the present low-energy model is not valid anymore, so that the present expression for the magnetic susceptibility can be trusted only at low temperatures.


\section{Equation of state and speed of sound}\label{EqState&SpeedSound}

Thanks to the BPS property of the magnetized layers, we can analytically derive the equation of state, $P=P(\epsilon)$, and subsequently determine the speed of sound, defined as $c_s ^2 =\partial P / \partial\epsilon$ (subject to the causality condition $0<c_s ^2 \leq 1$).

\begin{figure}[H]
    \centering
    \includegraphics[width=.8\linewidth]{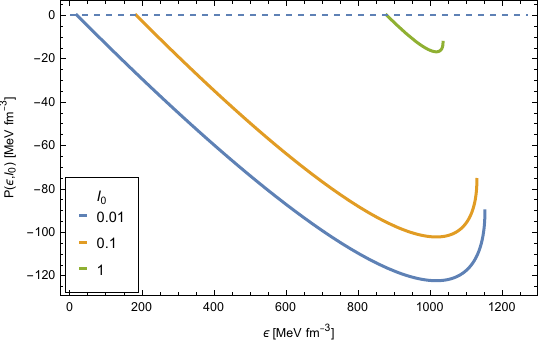}
    \caption{The Equation of state as a function of $\epsilon$ for $K=2\ \mbox{fm}^{-2}$, $L_r =25\ \mbox{fm}$.}
    \label{fig:EqState}
\end{figure}

To achieve this, let us consider the tensor ${T^\mu}_\nu=g^{\mu\rho}T_{\rho\nu}$, which possesses the following structure:
\begin{equation}
    {T^\mu}_\nu= \begin{pmatrix}
        -\epsilon &0&0&0\\
        0&T_1 &0&0\\
        0&0&T_2 &D\\
        0&0&D&T_3
    \end{pmatrix},
\end{equation}
where the energy density $\epsilon=T_{00}$ is defined in Eq. \eqref{EnDensityBPS}, and the remaining components are given by:
\begin{align}\notag
    T_1 &= -\frac{K \left(p^2 \cos ^2H+4 u^2 \sin ^2H\right)}{L^2}\\
    &\qquad\qquad\qquad\qquad+\frac{K(H')^2}{2L_r^2}+\frac{(u')^2}{L_r^2L^2},\\
    T_2 &= T_3=-\frac{K (H')^2}{2L_r^2},\\
    D &= \frac{K}{L^2}\left( p^2 \cos ^2H-4  u^2 \sin ^2H\right)-\frac{(u')^2}{L_r^2L^2}.
\end{align}
Upon diagonalization, the energy-momentum tensor assumes the form below, where the diagonal elements can be directly identified as the energy density and the principal normal stresses:
\begin{equation}
    {(T^{\text{D}})^\mu}_\nu= \begin{pmatrix}
        -\epsilon &0&0&0\\
        0&T_1 &0&0\\
        0&0&T_2+D &0\\
        0&0&0&T_2-D
    \end{pmatrix}.
\end{equation}
By incorporating the BPS relations $H=H(v)$, we can express the energy density $\epsilon = \epsilon(v)$ and the spatial eigenvalues of the energy-momentum tensor solely in terms of the variable $v$:
\begin{align}
    \epsilon &= 4K^2\left( 1+e^{-4v^2 -2I_0}(4v^2 -1) \right), \label{epsilonv}\\
    T_2+D&=-16K^2 v^2 e^{-4v^2 -2I_0},\\
    T_2-D &= T_1=0.
\end{align}
Assuming the system is a perfect fluid without dissipation, then the trace of the energy-momentum tensor takes the form ${(T^{\text{D}})^\mu}_\mu=-\rho+3P$. Therefore, we can associate the pressure to
\begin{align}\label{Pr}
    P=\frac{1}{3}(T_2+D)&=-\frac{16}{3}K^2 v^2 e^{-4v^2 -2I_0}.
\end{align}
Crucially, Eq. \eqref{epsilonv} allows for the inversion of the function to obtain $v(\epsilon)$ in a closed analytical form:
\begin{equation}\label{vEpsilon}
    v(\epsilon)=\pm\frac{1}{2}\sqrt{1-W_{0}\left( \left( 1-\frac{\epsilon}{4K^2} \right)e^{1+2I_0} \right)},
\end{equation}
where $W_0$ denotes the principal branch of the Lambert $W$ function. Substituting this result back into the pressure expression yields the analytic Equation of State ${P}(\epsilon)$:
\begin{equation}
    P(\epsilon)=\frac{4}{3}K^2 \left( 1-\frac{\epsilon}{4K^2}-e^{-1-2I_0 +W_{0}\left[ \left( 1-\frac{\epsilon}{4K^2} \right)e^{1+2I_0} \right] }  \right).
\end{equation}
The energy density $\epsilon$ is constrained by the square root in equation \eqref{vEpsilon}, taking values for which $W_0\left[ \left( 1-\frac{\epsilon}{4K^2} \right)e^{1+2I_0} \right] \leq 1$. This results in a negative pressure, in accordance with equation \eqref{Pr} (see Fig. \ref{fig:EqState}).

The endpoints of the plots are an artifact of the mathematical description, introduced by the Lambert function. Obviously, regions with negative compressibility are mechanically unstable and must be replaced by a phase coexistence construction; the corresponding branch of the equation of state is not physically realized \cite{Chomaz:Lambert}.

It is worth emphasizing that negative values for the pressure in the present case are not surprising since we are only considering states of pure baryonic
dense matter, neglecting the Coulomb interactions (and thus the electron gas surrounding the baryonic structures). Indeed, similar results were also obtained in \cite{Dorso:ColdNM,Alcain:LambdaElectron}, where (in the absence of a Coulomb potential) the presence
of non-homogeneous structures of pure nuclear matter is interpreted as a consequence of some periodic boundary conditions rather than the interplay between Coulomb and strong interactions. It should be noted that the application of periodic boundary condition is a fundamental characteristic of our model (as non-homogeneous Baryonic condensates do not appear if one adopts the usual
boundary conditions at spatial infinity).

It is worth mentioning that, when only the Coulomb interactions between nucleons are considered, stable nuclear pasta with positive pressure is permitted, even without the presence of the surrounding electron gas (see, for instance, \cite{Alcain:LambdaElectron}). While we did not investigate this scenario here, we will considered improving our analytic results including
Coulomb interactions in future work.

\begin{figure}[H]
    \centering
    \includegraphics[width=.8\linewidth]{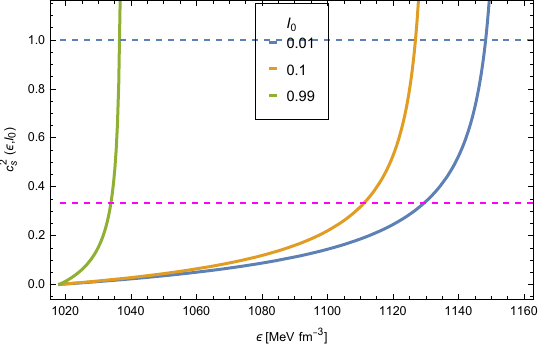}
    \caption{The speed of sound as a function of $\epsilon$ for $K=2\ \mbox{fm}^{-2}$, $L_r =25\ \mbox{fm}$. The magenta line represents the conformal limit $c_s^2=\frac{1}{3}$. The values of $\epsilon$ for which $c_s^2>1$ are not physical and are associated to the use of the Lambert function.}
    \label{fig:SpeedSound}
\end{figure}

Finally, the speed of sound is determined by differentiating the equation of state:
\begin{equation}
    c_s ^2 = \frac{\partial {P}}{\partial \epsilon}= \frac{1}{3}\left( \frac{1}{1+W_{0}\left[ \left( 1-\frac{\epsilon}{4K^2} \right)e^{1+2I_0} \right]} -1 \right).
\end{equation}
Once again, in order to obtain a positive value for $c_s ^2$, a constraint on the energy density needs to be imposed. In this case, $c_s ^2\geq 0$ when
\begin{align}
    W_{0}\left[ \left( 1-\frac{\epsilon}{4K^2} \right)e^{1+2I_0} \right]\leq 0.
\end{align}
This means that $\epsilon\geq 4K^2$. The values of the speed of sound have been represented in Fig. \ref{fig:SpeedSound}.

Here, there are two points that deserve attention.

In first place, one may notice that the speed of sound overcome the conformal limit of $\frac{1}{3}$. This behavior has been also observed in other model describing dense nuclear matter (see for instance \cite{Zhang:SpeedOfSound}). On the other hand, in the case studied in this paper the speed of sound overcome also the speed of light ($c_s^2=1$) and start to diverge in proximity of the end points of Fig. \ref{fig:EqState}. Once again, this is probably due to an artifact of the model. Ideally, the plots in Fig. \ref{fig:EqState} continue after the end points and the pressure becomes positive, undergoing a phase transition for high values of the energy density. In this situation, the speed of light does not diverge. Nevertheless, our model does not provide a description of the behavior of the system at such high energy densities.

The second point regards the values of the energy density itself. Indeed, the constraints imposed by the model result in energy densities that are very high with respect to the usual values obtained for nuclear pasta (or, more in general, for dense nuclear matter).

Beside these problems, the model allows for the definition of the Equation of States and to give an estimation of the speed of sound. It is generally a hard task to achieve for dense nuclear matter, especially through analytical tools.

\vspace{.5cm}
\section{The effects of isospin chemical potential: modified BPS bound and thermodynamics}\label{IsospinChemical}

In this section, we will include the effects of the isospin chemical potential. As it is well known, in the case of a field theory with a global SU(2) internal symmetry (see, for instance \cite{actor1985chemical, weldon1982covariant, loewe2006skyrme, ponciano2008skyrmions} and references therein), the effects of isospin chemical potential $\mu_{I}$ are introduced via the modified covariant derivative:

\begin{equation}
    D_{\mu}U \rightarrow\bar{D}_{\mu}U=D_{\mu}U+\mu_{I}[\tau_3,U]g_{\mu 0},
\end{equation}

Hence, due to the presence of the isospin chemical potential and still assuming the ansatz \eqref{UAnsatz} and \eqref{AAnsatz}, the energy density becomes the free energy density:

\begin{equation}\label{IEnergyDensity}
    T_{00}^I= T_{00}+ 2K\mu_{I}^{2} \sin^2 H,
\end{equation}
where $T_{00}$ is the ``old energy density" \eqref{EnDensity}. Obviously, the equations of motion are now obtained by minimizing the free energy density:

\begin{align}
    H'' + \left( \frac{L_r}{L}\right)^2\left(p^2-2L^2\mu_{I}^2-4u^2\right) \sin (2H)&=0,\\
     u''-4uKL_{r}^{2}\sin^2H&=0.
\end{align}

Quite remarkably, the system still admits a BPS reformulation. 
This can be seen by observing that the free energy density $T_{00}^{I}$ is a sum of squares plus a total derivative and a constant:
\begin{widetext}
\begin{equation}
    T_{00}^{I}= \frac{K}{2L_r^2} \left( H' + \frac{4L_r\sqrt{K+\mu_{I}^2}}{p}u\sin H \right)^2 
 + \frac{2(K+\mu_{I}^2)}{L_r^2p^2} \left( u' - \frac{pKL_r}{\sqrt{K+\mu_{I}^2}}\cos H \right)^2 
 + \frac{dW}{dr} + 2K\mu_{I}^{2}
\end{equation}
\end{widetext}
where the superpotential $W$ is:
\begin{equation}
    W=\frac{4K\sqrt{K+\mu_{I}^2}}{pL_r}u\cos{H},
\end{equation}
and the BPS condition now requires:
\begin{equation}
    L=\frac{p}{\sqrt{2(K+\mu_{I}^2)}} \leftrightarrow A=\pi^2\frac{p^2}{K+\mu_{I}^2}.
\end{equation}
Hence, the free energy reads
\begin{gather}
    F_{I}=\int d^3 x \sqrt{-g}T_{00}^{I}=AL_r \int_{0}^{2\pi}dr T_{00}^{I} \geq |E_I|, 
\end{gather}
where
\begin{align}
    E_I=\frac{AL_r}{e^2}|W(2\pi)-W(0)+4\pi K\mu_{I}^2| .
\end{align}
In this form, the modified BPS equations, which include the effects of the isospin chemical potential, are

\begin{align}\label{IBPSEq1}
    H' + \frac{4L_r\sqrt{K+\mu_{I}^2}}{p}u\sin H &=0,\\\label{IBPSEq2}
     u' - \frac{pKL_r}{\sqrt{K+\mu_{I}^2}}\cos H&=0.
\end{align}
Note that, as expected, the above isospin-dependent BPS equations imply the second-order field equations in the presence of a non-vanishing isospin chemical potential.
Moreover, in a similar way to the previous sections, the BPS condition with isospin chemical potential reduces to the following quadrature
\begin{equation}\label{int2}
    2\pi\sqrt{K}L_r=\int_{0}^{v(2\pi)}\frac{\sqrt{\frac{K+\mu_{I}^2}{K}}dv}{\sqrt{1-\exp\left( -\frac{4(K+\mu_{I}^2)}{K}v^2 -2I_0\right)}},
\end{equation}
with $p v(r)=u(r)$. Similarly to the previous sections we can write $v(2\pi)$ in terms of the magnetic flux in the $\varphi$ direction
\begin{equation}
    \Phi 
    = \frac{p^2\pi L_r }{\sqrt{2(K+\mu_{I}^2)}}(v(2\pi)-v(0)), 
\end{equation}
which implies that
\begin{align}
    v(2\pi)=\frac{\sqrt{2(K+\mu_{I}^2)}}{p^2\pi L_r}\Phi.
\end{align}

Note that due to the presence of the isospin chemical potential, the magnetic flux is reduced. In this way, we can rewrite the baryonic charge $B_I$ and the topological charge $E_I$ in terms of the magnetic flux and the isospin chemical potential
\begin{align}\label{IB}
    B_I &= \frac{4\sqrt{2(K+\mu_{I}^{2})}}{\pi L_r}\Phi \mathcal{M}_\Phi,\\ \label{IE}
    E_I &= \frac{4\sqrt{2}\pi K }{L_r}\Phi \sqrt{\mathcal{M}_\Phi} + \frac{4\pi^3 p^2L_r K \mu_{I}^{2}}{K+\mu_{I}^{2}},
\end{align}
where
\begin{align}
    \mathcal{M}_\Phi=1- \exp\left( -\frac{8(K+\mu_{I}^{2})^2}{Kp^4 \pi^2 L_r^2}\Phi^2 -2I_0 \right)
\end{align}
\\

\subsection{Useful Approximations and topological charge}

We can repeat the analysis in the previous sections to obtain the expression for the integration constant (see \eqref{int2}). Thus, it is useful to use the following rescaling
\begin{equation}
    s^2 = \frac{(K+\mu_{I}^{2})}{K}v^2\implies v=\sqrt{\frac{K}{K+\mu_{I}^{2}}}s.
\end{equation}

In this way, eq. \eqref{int2} is reduced to
\begin{equation}
    2\pi \sqrt{K}L_r = \int_{0}^{s(2\pi)} \frac{ds}{\sqrt{1-e^{-4s^2 -2I_0}}}
\end{equation}
which has the same form as \eqref{FVero}. Therefore, we can use the same approximations as in Section \ref{Approx} in order to find an analytical approximated expression of $v(2\pi)(I_0)$ 
and its inverse $I_0(v(2\pi))$:
\begin{widetext}
\begin{align}\label{ApproxSol2}
	s(2\pi)&=\sqrt {\frac{K+\mu_{I}^2}{K}}v(2\pi)\approx\left\{\begin{array}{l r}
	      \left[2\pi\sqrt{K}L_r+\frac{1}{4}\ln\left(1-e^{-2I_0}\right)\right]\qquad & \mbox{for }I_0\geq\varepsilon,\\
	      \sqrt {\frac {I_0}2} \sinh (4\pi \sqrt K L_r) & \mbox{for }I_0<\varepsilon,
	\end{array}\right.\\
\label{I0Approx2}
	I_0&\approx\left\{\begin{array}{l r}
	      -\frac{1}{2}\log\left[1-\exp\left(4\sqrt{\frac{K+\mu_{I}^{2}}{K}}v(2\pi)-8\pi\sqrt{K}L_r\right)\right]\qquad & \mbox{for }\sqrt{\frac{K+\mu_{I}^{2}}{K}}v(2\pi)\geq\tilde\varepsilon,\\
	      \frac{K+\mu_{I}^{2}}{K}v^2(2\pi)\frac{2}{\sinh^2 (4\pi \sqrt K L_r)} & \mbox{for }\sqrt{\frac{K+\mu_{I}^{2}}{K}}v(2\pi)<\tilde\varepsilon.
	\end{array}\right.
\end{align}
\end{widetext}

\begin{figure*}
    \centering
    \begin{subfigure}{0.3\textwidth}
        \includegraphics[width=\linewidth]{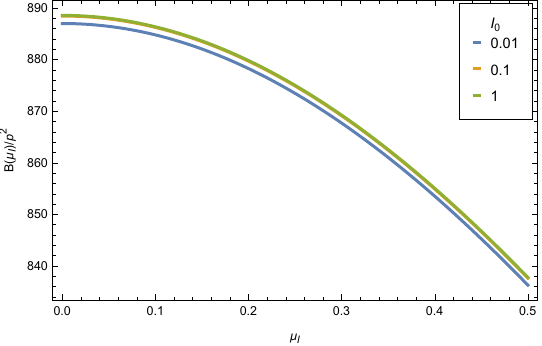}
        \caption{\label{Fig:BIMu}}
    \end{subfigure}\hspace{.3cm}
    \begin{subfigure}{0.3\textwidth}
        \includegraphics[width=\linewidth]{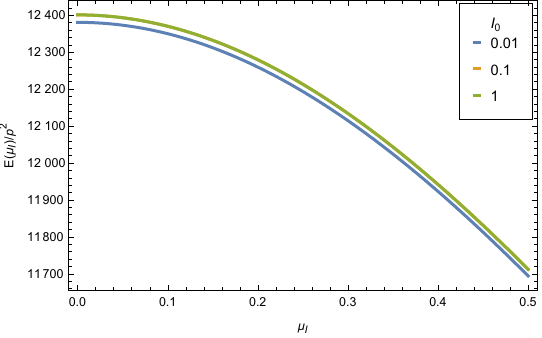}
        \caption{\label{Fig:EIMu}}
    \end{subfigure}\hspace{.3cm}
    \begin{subfigure}{0.3\textwidth}
        \includegraphics[width=\linewidth]{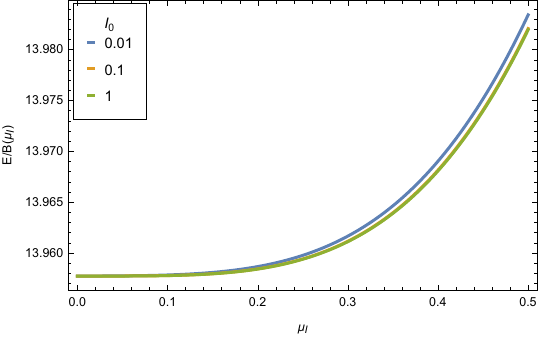}
        \caption{\label{Fig:EBIMu}}
    \end{subfigure}
    \caption{The baryonic charge (Fig \ref{Fig:BIMu}), the total energy (Fig \ref{Fig:EIMu}) and the energy per baryon (Fig \ref{Fig:EBIMu}) in terms of $\mu_I$, with $K=2\ \mbox{fm}^{-2}$, $L_r =25\ \mbox{fm}$.}
    \label{fig:EB}
\end{figure*}

This allows to define $B_I$ and $E_I$ in terms of $v(2\pi)$:
\begin{align}\label{BI}
    B_I&=4p^2v(2\pi)\ \mathcal{M},\\\label{EI}
    E_I &= \frac{4\pi^2 p^2 K}{\sqrt{K+\mu_{I}^{2}}}v(2\pi)\ \mathcal{M}+ \frac{4\pi^3 p^2L_r K \mu_{I}^{2}}{K+\mu_{I}^{2}},
\end{align}
where
\begin{align}\label{M}
    \mathcal{M}= 1-e^{-\frac{4(K+\mu_{I}^{2})}{K}v^2(2\pi)-2I_0}.
\end{align}
The baryonic charge and the energy, as well as the energy per baryons, are represented in Fig. \ref{fig:EB}.

\subsection{The partition function and its dependence on the isospin}

As for the previous case, we can consider call
\begin{align}\label{IBInteger}
    n=\frac{B_I}{p^2},
\end{align}

where $n$ is considered as a continuous quantity, motivated by the observations outlined in the previous section. 

As for the case with $\mu_I = 0$, the quantity $n$ converges to a finite value, which deoends on $\mu_I$ as follows

\begin{equation}
    n_{\textit{max}}(\mu_I)= \lim_{\Phi \rightarrow \Phi_{max}}\frac{B_I}{p^2}=\frac{8\pi K L_r }{\sqrt{K+\mu_{I}^2}}.
\end{equation}

Notice that the existence of a maximum for magnetic flux can be directly derived from the approximation for $I_0$ in equation \eqref{I0Approx2}:

\begin{equation}
    \Phi_{max}= \frac{\sqrt{2}\pi^2 p^2 KL_{r}^{2}}{K+\mu_{I}^2}.
\end{equation}

\begin{figure}[h!]
    \centering
    \includegraphics[width=7cm]{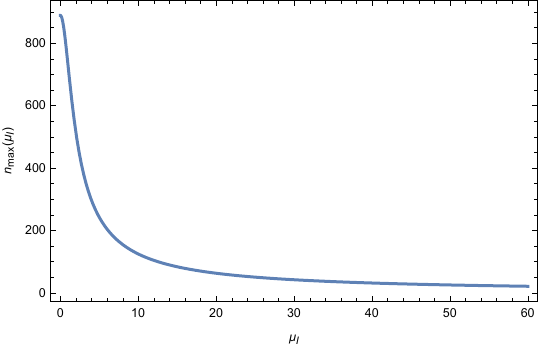}
    \caption{The quantity $n_{max}$ in terms of $\mu_I$, where $K=2\ \mbox{fm}^{-2}$, $L_r =25\ \mbox{fm}$.}
    \label{fig:nMax}
\end{figure}

It is worth emphasizing here that the presence of the isospin chemical potential decreases the maximum value for $n$ (see Fig. \ref{fig:nMax}).
With this in mind, we can compute the grand canonical partition function as follows 

\begin{equation}\label{IPartitionFunction}
    \mathcal{Z}(\beta, \mu_B, \mu_I)= \iint\sum_{p=-\infty}^{+\infty}e^{-\beta p^2\mathcal{F}_I (n,\mu_B, \mu_I)}\ dn\ dn_I.
\end{equation}
where 
\begin{equation}\label{IFreeEnergy}
    \mathcal{F}_I(n, \mu_B, \mu_I)= G_I-n\mu_B-n_I\mu_I,
\end{equation}
and $n_I$ represents the \textit{isospin number}. We will see later that the quantity $n_I$ is not independent, but rather depends on $n$. Therefore, the double integral in the definition of the partition function \eqref{IPartitionFunction} can be reduced to a single integral over $n\in(0,n_{max}(\mu_I)]$. 

Using equations \eqref{IB} and \eqref{IE}, 
\begin{align}\label{GI}
    G_I=\frac{2\pi^2 K}{\sqrt{K+\mu_{I}^{2}}}\sqrt{nv(2\pi)(n,\mu_I)}+\frac{4\pi^3  K L_r \mu_{I}^{2}}{K+\mu_{I}^{2}}.
\end{align}

\begin{figure}[h!]
    \centering
    \begin{subfigure}{0.4\textwidth}
        \includegraphics[width=\linewidth]{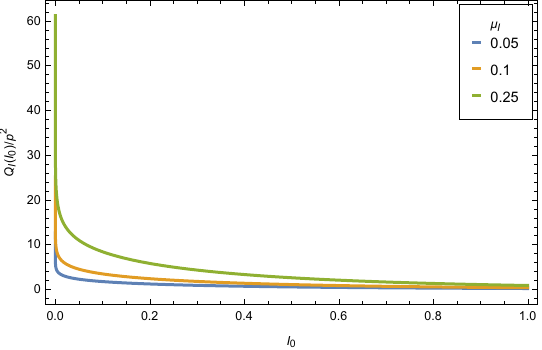}
        \caption{\label{Fig:QII0}}
    \end{subfigure}\vspace{.5cm}
    \begin{subfigure}{0.4\textwidth}
        \includegraphics[width=\linewidth]{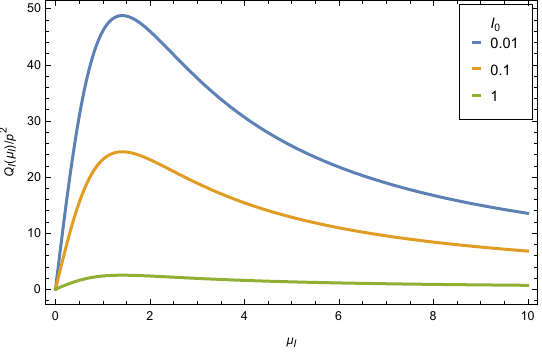}
        \caption{\label{Fig:QIMu}}
    \end{subfigure}
    \caption{Here is represented the behavior of $Q_I/p^2$ as a function of $I_0$ for different values of $\mu_I$ (Fig. \ref{Fig:QII0}) and as a function of $\mu_I$ (Fig. \ref{Fig:QIMu}). As usual, $K=2\ \mbox{fm}^{-2}$, $L_r =25\ \mbox{fm}$. It is noteworthy that $Q_I$ has a maximum for a specific value of $\mu_I$ and then it goes to zero for big values of the latter. This is due to the decreasing in total number of baryons, as depicted in the first graph of Fig. \ref{fig:EB}}.
    \label{fig:QI}
\end{figure}

In order to understand the behavior of the partition function, we need to analyze the free energy in more details. In particular, the quantity we have called $n_I$ corresponds to the isospin charge defined as
\begin{align}
    Q_I=\int d^3x J_0^3,
\end{align}
where $J_0^3$ is the component with $\mu=0$ along $\tau_3$ of the isospin current
\begin{align}
    J_\mu^a=2K\mbox{Tr}\left(D_\mu L L^{-1} \tau^a\right),
\end{align}
and $D_\mu$ is the usual covariant derivative. This gives
\begin{align}
    Q_I=\frac{8KL_r  \mu_I \pi^2 p^2}{K+\mu_{I}^2}\int dr \sin^2H,
\end{align}
which is the volume integral of the isospin component of the energy density \eqref{IEnergyDensity}. By using the BPS equations, we can rewrite the isospin charge in terms of $v(r)$ as follows 
\begin{align}\label{QI}
    Q_I = \frac{8\mu_I \sqrt{K} \pi^2 p^2 }{{K+\mu_{I}^{2}}}\ \mathcal{I},
\end{align}
where
\begin{align}\label{I}
    \mathcal{I}=\int_{0}^{\sqrt{\frac{K+\mu_{I}^{2}}{K}}v(2\pi)} \frac{e^{-4x^2-2I_0}}{\sqrt{1-e^{-4x^2-2I_0}}}dx.
\end{align} 
The numerical solution to this integral is graphically represented in Fig. \ref{fig:QI}.

As just anticipated, $Q_I$, and thus $n_I$, is not an independent quantity, but it depends on the baryonic charge. In particular, one can use the relation \eqref{IB} in order to write $I_0$ in terms of $n$. Then, the obtained expression can be used in equation \eqref{BI}. This way, the sum in the partition function reduces to the sum over $n$.

\begin{figure}[H]
    \centering
    \includegraphics[width=0.4\textwidth]{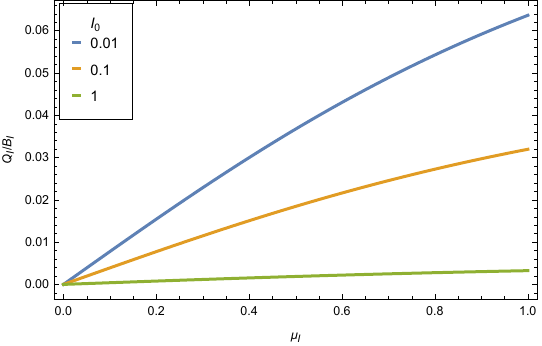}
    \caption{The ratio $Q_I/B_I$ as a function of $\mu_I$, with $K=2\ \mbox{fm}^{-2}$, $L_r =25\ \mbox{fm}$.}
    \label{fig:QI/BI}
\end{figure}

Let us observe that the ratio of the isospin number and the baryonic number is determined by 
\begin{eqnarray}\label{Q/I}
    \frac{Q_{I}}{B_{I}} =\frac{2\pi^2\sqrt{K} \mu_{I}}{{(K+\mu_{I}^{2})}v(2\pi)\ \mathcal{M}}\ \mathcal{I},
\end{eqnarray}
where $\mathcal{M}$ and $\mathcal{I}$ has been defined in \eqref{M} and \eqref{I} respectively. In Fig. \ref{fig:QI/BI} is represented a numerical solution of this quantity. It can be observed that, when $I_0$ is fixed, and thus fixing the boundary conditions on the solution $H(r)$, the quantity $Q_I/B_I$ becomes constant for big values of the isospin chemical potential.

A good approximation for the integral \eqref{I} can be obtained using the line of Appendix \ref{App:Approximation}. When $I_0$ is not too small (i.e., $I_0>\epsilon$), the approximation reads

\begin{align}\label{IntApprox}
   \mathcal{I}\approx\frac{1}{8}\left[  Li_2\left(e^{-2I_0}\right)-2\log{\left(1-e^{-2I_0}\right)}\right].
\end{align} 
Furthermore, if $I_0\leq\epsilon$ (meaning that also $\C v^2(2\pi)$ is small, as outlined in equation \eqref{ApproxSol2}) and calling ${x=\C\frac{4v^2(2\pi)}{I_0}}$, then
\begin{align}
    \mathcal{I}&\approx\frac{1}{4}\left[\ln\left(1+x+\sqrt{x(2+x)}\right)\right.\\ \notag
    &\quad-\left.\frac{\frac{K+\mu_I^2}{K}v^2(2\pi)}{x}\left(\sqrt{(1+x)x}+\mbox{arcsin}(\sqrt{x})\right)\right].
\end{align}

\begin{figure}[H]
    \centering
\includegraphics[width=0.4\textwidth]{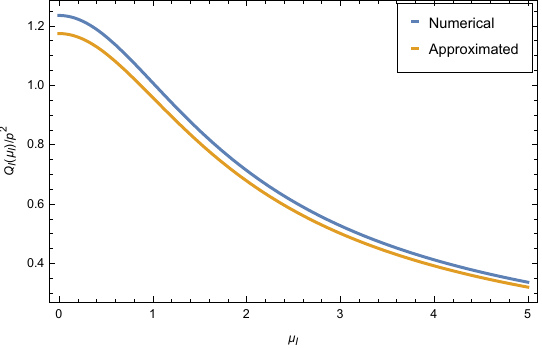}
  \caption{Approximation of the integral \eqref{I} and its numerical values as a function of $\mu_I$, with  $K=2\ \mbox{fm}^{-2}$, $L_r =25\ \mbox{fm}$ and $I_0=0.1$. $v(2\pi)$ given by equation \eqref{ApproxSol2}. }
    \label{fig:ApproxIntegral}
\end{figure}

The comparison of this approximation and the numerical integration is represented in Fig. \ref{fig:ApproxIntegral}.

Thus, we can pose
\begin{equation}
  \alpha(n,\mu_I)=\frac{Q_{I}}{B_{I}}= \frac{\pi^2\mu_I}{4v(2\pi)(n,\mu_I)}\frac{\sqrt{K}}{K+\mu_I^2}\ \hat{\mathcal{I}}(n,\mu_I),
\end{equation}
where
\begin{align}
    \hat{\mathcal{I}}(n,\mu_I)= \frac{A+\C v(2\pi)^2(n,\mu_I) B
    }{\C v(2\pi)^2(n,\mu_I)\left(1+\frac{1}{\sinh^2(4\pi\sqrt{K}L_r)}\right)}
\end{align}
for $I_0\approx 0$, where
\begin{align}
    A&=\frac{1}{2}\ln\left[\cosh^2(4\pi\sqrt{K}L_r)\right.\\ \notag
    &+\left.\sinh(4\pi\sqrt{K}L_r)\sqrt{1+\cosh^2(4\pi\sqrt{K}L_r)}\right],\\
    B&=-\frac{\cosh(4\pi\sqrt{K}L_r)}{\sinh(4\pi\sqrt{K}L_r)}-\frac{4\pi\sqrt{K}L_r}{\sinh^2(4\pi\sqrt{K}L_r)},
\end{align}
and
\begin{align}
    \hat{\mathcal{I}}(n,\mu_I)= \frac{  Li_2\left(e^{-2I_0(n,\mu_I)}\right)-2\log{\left(1-e^{-2I_0(n,\mu_I)}\right)}}{1-e^{-4\frac{K+\mu_I^2}{K}v(2\pi)^2(n,\mu_I)-2I_0(n,\mu_I)}}
\end{align}
for $I_0>\epsilon$.

The dependence of $v$ and $I_0$ from $n$ and $\mu_I$ has been defined by using the relations \eqref{ApproxSol2}, \eqref{I0Approx2} and imposing the condition \eqref{IBInteger}. The free energy reads
\begin{equation}\label{IFreeEnergy}
    \mathcal{F}_I(n, \mu_B, \mu_I)= n[\mu^*(n,\mu_I)-\mu_B-\alpha(n,\mu_I)\mu_I)],
\end{equation}
where $\mu^*(\mu_I,n)=G_I(\mu_I,n)/n$. 

\begin{figure}[h!]
    \centering
    \begin{subfigure}{0.4\textwidth}
        \includegraphics[width=\textwidth]{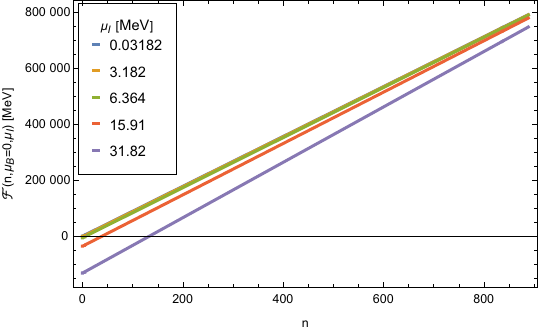}
        \caption{\label{Fig:Fn}}
    \end{subfigure}
    
    \vspace{.5cm}
    \begin{subfigure}{0.4\textwidth}
        \includegraphics[width=\textwidth]{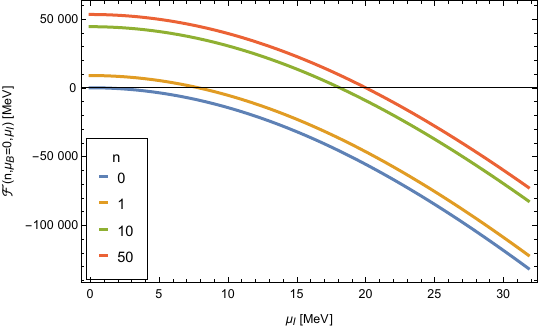}
        \caption{\label{Fig:FmuI}}
    \end{subfigure}
    
    \caption{Representation of the free energy as functions of $n$ and $\mu_I$. Here, $K=2\ \mbox{fm}^{-2}$, $L_r =25\ \mbox{fm}$ and $\mu_B=0$.}
    \label{fig:FunctionsMu}
\end{figure}

As for the case with $\mu_I=0$, in order to have a well defined partition function, the free energy needs to be positive. It is possible to study its sign in terms of $\hat{n}$, $\mu_I$ and $\mu_B$ through the following observations.

We start by considering $\mu_B=0$, remembering that the approximation previously discussed is valid for two different ranges of $I_0$, and thus two different ranges of $\hat{n}=\sqrt{\C}n$ (see equations \eqref{ApproxSol2} and \ref{I0Approx2}). For this reason, we should discuss the behavior of free energy in terms of $(\hat{n},\mu_I)$, instead of $(n,\mu_I)$. With this in mind, the behavior of the free energy in terms of both $\hat{n}$ and $\mu_I$ is represented in Fig. \ref{fig:FunctionsMu}. In particular, the plot in Fig. \ref{Fig:Fn} shows the dependence of $\mathcal{F}(\hat{n},0,\mu_I)$ from $\hat{n}$, which is mostly linear. For this reason, we can consider the linear approximation as for the case with $\mu_I=0$. In this limit, the expression of the free energy becomes
\begin{align}
    \mathcal{F}(\hat{n},0,\mu_I)&\approx\frac{\pi^2\sqrt{K}}{K+\mu_I^2}\left(K+3\mu_I^2\right)\hat{n}\\ \notag
    &\qquad-\frac{20\pi^3\mu_I^2KL_r}{{K+\mu_I^2}}.
\end{align}

Both the plots in \ref{fig:FunctionsMu} reveal that for some values of $(\hat{n},\mu_I)$ the free energy is negative. Through the linear approximation is easy to find out that, in order to avoid this issue, $\hat n$ needs to satisfy
\begin{align}
    \hat{n}\geq\frac{\mu_I^2}{K+3\mu_I^2}20\pi\sqrt{K}L_r.
\end{align}
This means that for each $\mu_I$ there exists a lower bound for $\hat{n}$. This lower bound is zero only when $\mu_I=0$. This situation is represented in the plots of Fig. \ref{fig:FunctionsMu}.

Considering the analysis outlined above, when $\mu_B=0$ the partition function takes the form
\begin{align}
    \mathcal{Z}(\beta, 0, \mu_I)=\sqrt{\frac{K}{K+\mu_I^2}}\int_{\hat{n}_{min}(\mu_I)}^{\hat{n}_{max}}\xi\left(\frac{i\beta}{\pi}\mathcal{F}(\hat{n},0, \mu_I)\right)\ d\hat{n},
\end{align}
where $\hat{n}_{min}(\mu_I)=\frac{\mu_I^2}{K+2\mu_I^2}(1+12\sqrt{K}L_r)$, $\hat{n}_{max}=8\pi\sqrt{K}L_r$.

When also $\mu_B$ is considered, then
\begin{align}\label{FLinMuB}\notag
    \mathcal{F}(\hat{n},\mu_B,\mu_I)&\approx\frac{\pi^2\sqrt{K}}{K+\mu_I^2}\left[\left(K+3\mu_I^2-\frac{\sqrt{K+\mu_I^2}}{\pi^2}\mu_B\right)\hat{n}\right.\\ 
    &\qquad-\left.20\mu_I^2\pi\sqrt{K}L_r\right].
\end{align}
Therefore, the presence of the baryonic chemical potential also affects the range of $n$, giving 
\begin{align}
    \hat{n}_{min}(\mu_B,\mu_I)=\frac{20\pi\sqrt{K}L_r\mu_I^2}{K+3\mu_I^2-\frac{\sqrt{K+\mu_I^2}}{\pi^2}\mu_B}
\end{align}
This defines an upper bound for $\mu_B$. Indeed, it is evident from \eqref{FLinMuB} that $\mu_B$ has to satisfy
\begin{align}
    \mu_B\leq\pi^2\frac{K+3\mu_I^2}{\sqrt{K+\mu_I^2}}\equiv\tilde\mu_B(\mu_I).
\end{align}
Furthermore, another necessary condition is $\hat{n}_{max}-\hat{n}_{min}(\mu_B,\mu_I)\geq 0$. This is obtained when
\begin{align}
    \mu_B\leq\tilde\mu_B(\mu_I)-\frac{5\mu_I^2\pi^2}{2\sqrt{K+\mu_I^2}},
\end{align}
which is smaller than $\tilde\mu_B(\mu_I)$. When $\hat{n}_{min}(\mu_B,\mu_I)$ approaches $\hat{n}_{max}$, then the partition function approaches to zero.

\begin{figure}
    \centering
    \begin{subfigure}{0.4\textwidth}
        \includegraphics[width=\textwidth]{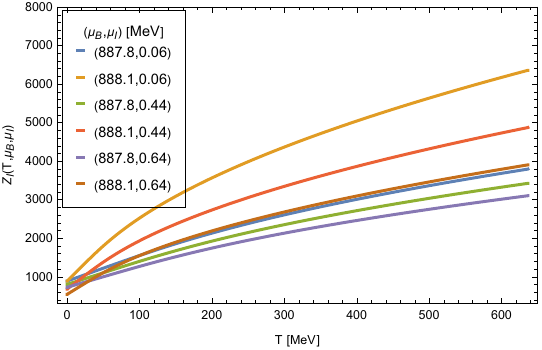}
        \caption{\label{Fig:PFMuISmall}}
    \end{subfigure}

    \vspace{0.3cm}
    \begin{subfigure}{0.4\textwidth}
        \includegraphics[width=\textwidth]{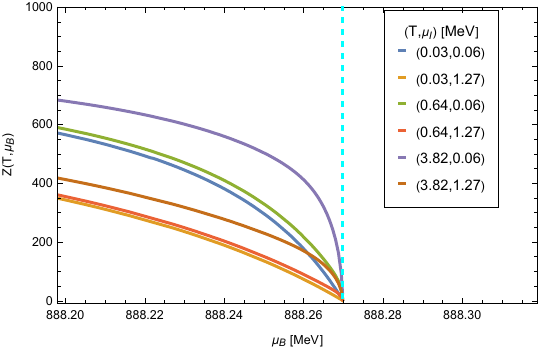}
        \caption{\label{Fig:PFMuIBig} }
    \end{subfigure}
    \caption{The partition function in terms of $T$ and $\mu_B$ for $K=2\ \mbox{fm}^{-2}$, $L_r =25\ \mbox{fm}$.}
    \label{fig:PFMuB}
\end{figure}

A representation of the partition function in terms of $T$ and $\mu_B$, for different values of $\mu_I$, is given in Fig. \ref{fig:PFMuB}.

The meaning of this behavior does not seem to have an immediate interpretation. Nevertheless, we need to remember that the analytical description of the isospin within strongly coupled systems of baryonic crystals is generally an hard task. The study represents a suggestion for a possible direction.

\subsection{Thermodynamical quantities and some observations}
{In this section, we analyze the same thermodynamic quantities considered in the previous one, but in the presence of the isospin. These quantities are represented in figures \eqref{fig:AvarageParticlesI} and \ref{fig:IIE}.
It is straightforward to verify that for $\mu_I = 0$ the same expression for the free energy is recovered; therefore, we shall focus only on the cases with $\mu_I \neq 0$. 

Observing the average number of particles in Fig. \eqref{fig:AvarageParticlesI}, it appears evident that the necessary temperature for particles production increases with $\mu_I$. On the other hand, as also observed in the previous section with $\mu_I=0$, for increasing values of $\mu_B$, the particle production becomes possible also for small $T$. 

The behavior of the rest of the thermodynamical quantities in therms of $\mu_I$ is quite interesting. Indeed, we can observe that the nuclear matter starts to condensate as $\mu_I$ increases. As depicted in figures \eqref{Fig:IHCTMu} and \eqref{Fig:IIETMu}, both the heat capacity and internal energy decrease dramatically, reaching negative and non physical values. It could indicates the presence of a phase transition that our model is not able to describe.

Finally, one can observe that once again the plot of the entropy in Fig. \ref{fig:IIE} shows the same behavior as for $\mu_I=0$. Namely, the entropy is not zero for $T\rightarrow 0$. In this case, the limit value of the entropy depends on $\mu_I$.
\vspace{1cm}
\begin{figure}[htb!]\centering

    \vspace{0.5cm}
        \textbf{Average number of particles}
    \vspace{0.5cm}
    
    \begin{subfigure}{0.4\textwidth}        
        \includegraphics[width=\linewidth]{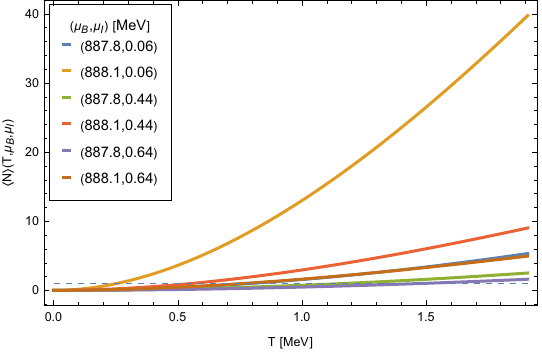}
        \caption{\label{Fig:INPT}}
    \end{subfigure}\vspace{.3cm}
    
    \begin{subfigure}{0.4\textwidth}
        \includegraphics[width=\linewidth]{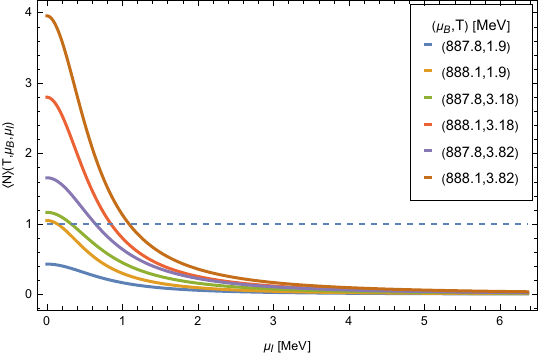}
        \caption{\label{Fig:IHCT}}
    \end{subfigure}
    \caption{Average number of particle in terms of $T$ and $\mu_I$ for $K=2\ \mbox{fm}^{-2}$, $L_r =25\ \mbox{fm}$.}
\label{fig:AvarageParticlesI}
\end{figure}

\begin{figure*}[htb!]\centering
    \vspace{2cm}
        \textbf{Heat capacity}
    \vspace{0.5cm}
    
    \begin{subfigure}{0.4\textwidth}
        \includegraphics[width=\linewidth]{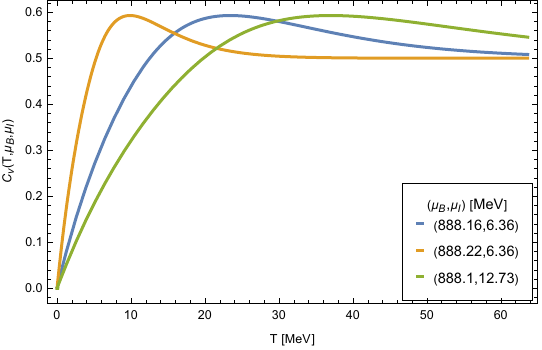}
        \caption{\label{Fig:IHCT}}
    \end{subfigure}\hspace{.3cm}
    \begin{subfigure}{0.4\textwidth}
        \includegraphics[width=\linewidth]{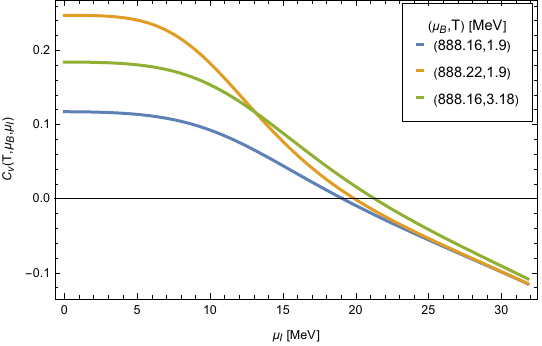}
        \caption{\label{Fig:IHCTMu}}
    \end{subfigure}\hspace{.3cm}

    \vspace{.5cm}
        \textbf{Internal energy}
    \vspace{0.5cm}
    
    \begin{subfigure}{0.4\textwidth}        
        \includegraphics[width=\linewidth]{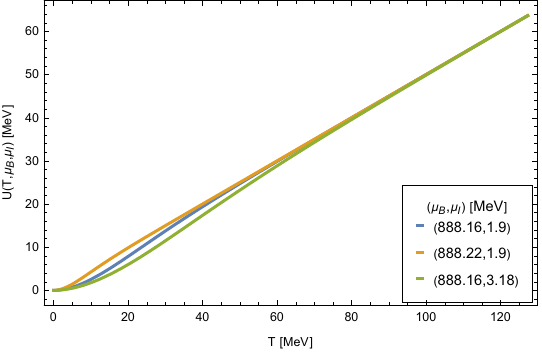}
        \caption{\label{Fig:IIET}}
    \end{subfigure}\hspace{.3cm}
    \begin{subfigure}{0.4\textwidth}        
        \includegraphics[width=\linewidth]{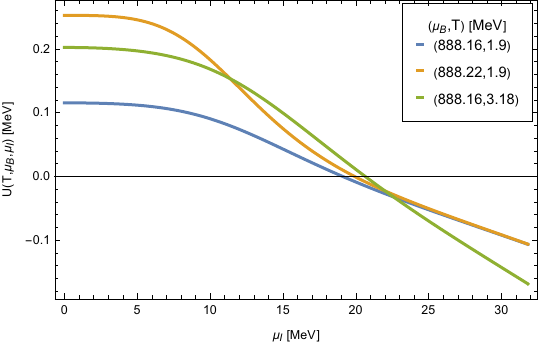}
        \caption{\label{Fig:IIETMu}}
    \end{subfigure}\hspace{.3cm}

    \vspace{.5cm}
        \textbf{Entropy}
    \vspace{0.5cm}
        
    \begin{subfigure}{0.4\textwidth}
        \includegraphics[width=\linewidth]{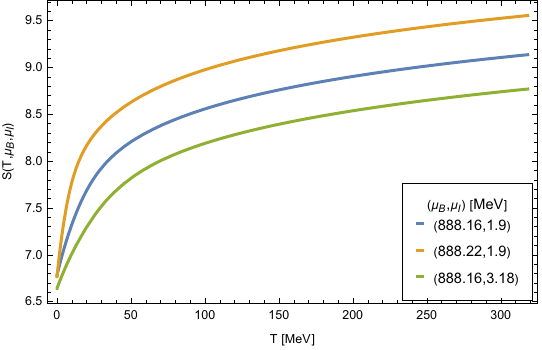}
        \caption{\label{Fig:IST}}
    \end{subfigure}\hspace{.3cm}
    \begin{subfigure}{0.4\textwidth}
        \includegraphics[width=\linewidth]{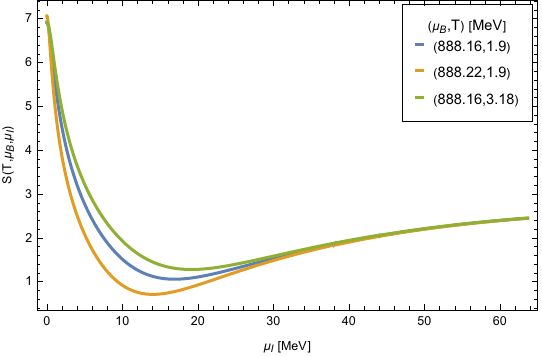}
        \caption{\label{Fig:ISTMu}}
    \end{subfigure}\hspace{.3cm}
\caption{Thermodynamical quantities as functions of $T$ and $\mu_I$ for $K=2\ \mbox{fm}^{-2}$, $L_r =25\ \mbox{fm}$.}
\vspace{1cm}
\label{fig:IIE}
\end{figure*}

\vspace{-.9cm}

\subsection{Equation of State and speed of sound with non-zero isospin chemical potential}

 The coupling between the G-NLSM and the isospin chemical potential, subject to the BPS conditions for magnetized layers, slightly modifies the structure of the energy-momentum tensor, which becomes 
 \begin{equation}
    {T^\mu}_\nu= \begin{pmatrix}
        -\epsilon &0&-L^2C&L^2C\\
        0&T_1 &0&0\\
        C&0&T_2 &D\\
        -C&0&D&T_3    
    \end{pmatrix},
\end{equation}
where 
\begin{align}\notag
    \epsilon &= 2K\mu_I^2\sin^2H+\frac{K}{L^2}\left[p^2\cos^2(H)+4\sin^2(H)u^2\right]\\
    &\qquad\qquad\qquad\qquad+\frac{K(H')^2}{2L_{r}^{2}}+\frac{(u')^2}{(L_r L)^2}
\end{align}
\begin{align}\notag
    T_1 &= -2K\mu_I^2\sin^2H-\frac{K \left(p^2 \cos ^2H+4 u^2 \sin ^2H\right)}{L^2}\\
    &\qquad\qquad\qquad\qquad+\frac{K(H')^2}{2L_r^2}+\frac{(u')^2}{L_r^2L^2},
\end{align}
\begin{align}
    T_2 &= T_3=-2K\mu_I^2\sin^2H-\frac{K (H')^2}{2L_r^2},
\end{align}
\begin{align}
    D &=\frac{K}{L^2}\left( p^2 \cos ^2H-4  u^2 \sin ^2H\right)-\frac{(u')^2}{L_r^2L^2},
\end{align}
\begin{align}
    C&=\frac{4K\mu_I}{L^2}u\sin^2H.
\end{align}
Using the BPS conditions and after the diagonalization, the matrix reads
\begin{equation}
    {(T^{\text{D}})^\mu}_\nu= \begin{pmatrix}
        -\frac{\epsilon}{2}+\sqrt{\Omega} &0&0&0\\
        0&0 &0&0\\
        0&0&T_2+D &0\\
        0&0&0&-\frac{\epsilon}{2}-\sqrt{\Omega}
    \end{pmatrix},
\end{equation}
where $\Omega=\frac{\epsilon^2}{4}-2L^2C^2$. Finally, using the trace of energy- momentum tensor, the pressure can be associated to ${P=\frac{1}{3}(T_2+D)}$. Both the energy density $\epsilon$ and the pressure $P$ can be expressed as functions of the profile  $v(r)$ as follows
\begin{align}\notag
    \epsilon(v,\mu_I) &= 4K^2\left[ 1+e^{-\frac{4(K+\mu_I^2)}{K}v^2 -2I_0}\left(\frac{4(K+\mu_I^2)}{K}v^2 -1\right) \right]\\ \label{epsilonvmuI}
    &\qquad +2K\mu_I^2,\\
    P(v,\mu_I) &= \frac{2 K^2}{3} \left( \frac{\mu_I^2}{K+\mu_I^2} - 8 v^2 e^{ - \frac{4(K+\mu_I^2) v^2}{K}-2I_0} \right).
\end{align}

Also in this case, the energy density can be inverted, giving
\begin{align}\label{vEpsilon}
    v(\epsilon)&=\pm\sqrt{\frac{K}{4(k+\mu_I^2)}}\\ \notag
    &\quad\times\sqrt{1-W_{0}\left[ \left( 1+\frac{\mu_I^2}{2K}-\frac{\epsilon}{4K^2} \right)e^{1+2I_0} \right]}.
\end{align}
Therefore, the energy density can be written in terms of  $\epsilon$ as
\begin{align}\notag
    P(\epsilon)&=\frac{4}{3}K^2 \left( 1+\frac{\mu_I^2}{2K}-\frac{\epsilon}{4K^2} \right)\\ \label{PmuI}
    &\quad-\frac{4}{3}K^2e^{-1-2I_0 +W_{0}\left[ \left( 1+\frac{\mu_I^2}{2K}-\frac{\epsilon}{4K^2} \right)e^{1+2I_0} \right] } .
\end{align}

\vspace{.5cm}
The representation of the pressure in terms of the energy density for different values of $\mu_I$ is given in Fig. \ref{fig:ISOEoS}.

\begin{figure}[H]
    \centering
    \includegraphics[width=.9\linewidth]{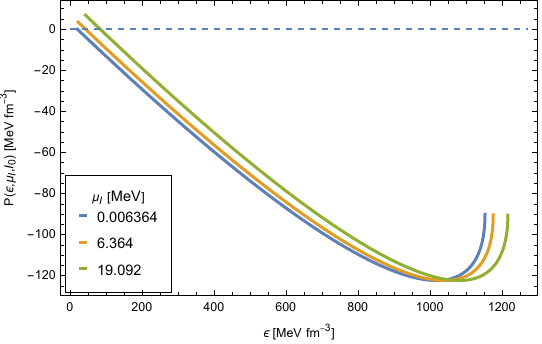}
    \caption{Plot of the pressure for $K=2\ \mbox{fm}^{-2}$, $L_r =25\ \mbox{fm}$, $I_0=0.01$ for different values of $\mu_I$. }
    \label{fig:ISOEoS}
\end{figure}
The speed of sound $c_s^2$, finally, results
\begin{equation}
c_s^2 = \frac{\partial {P}}{\partial \epsilon}= \frac{1}{3}\left( \frac{1}{1+W_{0}\left[ \left( 1+\frac{\mu_I^2}{2K}-\frac{\epsilon}{4K^2} \right)e^{1+2I_0} \right]} -1 \right),
\end{equation}

\clearpage
which is represented in Fig. \ref{fig:SpeedSoundMuI}. 
\begin{figure}[h]
    \centering
    \includegraphics[width=.9\linewidth]{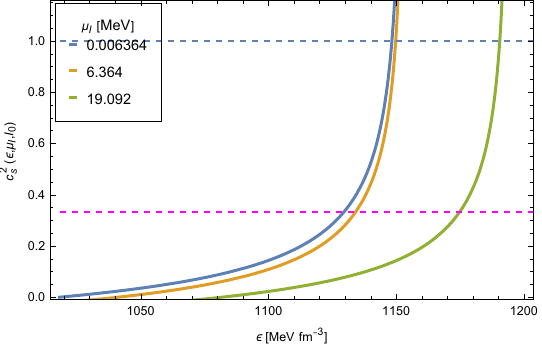}
    \caption{The speed of sound as a function of $\epsilon$ for $K=2\ \mbox{fm}^{-2}$, $L_r =25\ \mbox{fm}$, $I_0=0.01$ for different values of $\mu_I$. The magenta line represents the the conformal limit $c_s^2=\frac{1}{3}$. The values of $\epsilon$ for which $c_s^2>1$ are not physical and are associated to the use of the Lambert function.}
    \label{fig:SpeedSoundMuI}
\end{figure}
In the limit $\mu_I \to 0$, the expressions for $\epsilon(v)$ and $P(v)$ consistently recover the results presented in the previous section.

The observation reported in Section \ref{EqState&SpeedSound} for the case with $\mu_I=0$ are valid also in this case.

\section{Conclusions and
perspectives}\label{Conclusion}

In the present manuscript (using techniques from the Hamilton-Jacobi equation of classical mechanics, from the theory of BPS solitons as well as from the theory of the Casimir effect), we have analyzed the semi-classical thermodynamics of BPS magnetized baryonic layers (possessing both baryonic charge and magnetic flux) in the gauged non-linear sigma model (G-NLSM) minimally coupled to Maxwell theory. Such a theory describes the low-energy limit of QCD together with the electromagnetic interactions of the hadronic degrees of freedom. In the case in which the isospin chemical potential vanishes, the topological charge that naturally appears on the right-hand side of the BPS bound (and which plays the role of the energy of the baryonic layers) is a non-linear function of the baryonic charge. Correspondingly, the thermodynamics of these magnetized baryonic layers is very non-trivial. We have computed analytically the relations between baryonic charge, topological charge, magnetic flux, and relevant thermodynamical quantities (such as pressure, specific heat, and magnetic susceptibility) of these layers. We have determined the critical baryonic chemical potential beyond which the model cannot be trusted anymore. Quite remarkably, this analysis can also be extended to include the effects of the isospin chemical potential. Indeed, the isospin chemical potential is essential in the description of the nuclear pasta phase. We have explicitly constructed the BPS bound and the corresponding BPS configurations in the case in which the isospin chemical potential is non-zero. The topological charge that naturally appears on the right-hand side of the BPS bound plays the role of the free energy of the baryonic layers. Also, in this case, this charge is a non-linear function of the baryonic charge. On the technical side, it is interesting to note that the grand canonical partition function can be related to the Riemann zeta function. It is quite a non-trivial achievement to derive explicit expressions valid at finite baryon density for all these thermodynamic quantities of a strongly interacting magnetized system at finite baryon density (especially taking into account the problems of lattice QCD at finite baryon density). The properties of these thermodynamical quantities (which also include the equation of state and the speed of sound) appear to be very reasonable and can be useful to improve our analytic understanding of strongly magnetized baryonic systems as well as in comparison with lattice results.

A further and final comment here is in order. We mentioned in the introduction the possibility of describing nuclear pasta states by means of our solutions (in particular, \textit{lasagna-like structures}). Nevertheless, a direct comparison between the physical quantities obtained from our model and the physics of neutron stars is not possible at this stage of the work, and would require a further extension of the model. Indeed, we are able to describe the physical properties of ordered structures of baryons organized in layers, reminding of the solid part of \textit{lasagna} structures; this achievement is significant because such results are generally difficult to obtain using analytical tools. However, our analysis misses the contribution of the Coulomb forces, the liquid/gaseous nuclear matter and the electron gas surrounding the baryonic layers, an omission that is fundamental to address for a reliable comparison with the physical data. This extension is the subject for a future work.

\section*{Acknowledgments}
F.C. has been funded by FONDECYT Grant No. 1240048 and by Grant ANID EXPLORACI\'ON 13250014.

L.R. acknowledges the Ministero dell'Universit\`a e della Ricerca (MUR), PRIN2022 program (Grant PANTHEON 2022E2J4RK) for partial support.

F.M. acknowledges the support of the Next Generation EU - Prin 2022 project "Singular Interactions and Effective Models in Mathematical Physics- 2022CHELC7".

\begin{appendix}
 \section{On the approximation of the integral \eqref{equation}} \label{App:Approximation}
Equation \eqref{equation} is assumed to be independent of $p$ and relates $v=u(2\pi)/p$ to $I_0$. $I_0$ is thus expected to be independent of $p$. To solve this equation one should be able to compute the integral \eqref{FVero} that we rewrite as
 \begin{align}
 G(v,I)&=\int_0^v \frac {dx}{\sqrt{1-e^{-(4x^2+2I)}}}\cr
 &=v+K(v,I).
 \end{align}
 where
 \begin{align}
     K(v,I)=\int_0^v \Big(\frac {1}{\sqrt{1-e^{-(4x^2+2I)}}}-1\Big)dx.
 \end{align}
 Now, let us consider the function $K$. Its derivative w.r.t. $v$ is
 \begin{align}
 \frac {1}{\sqrt{1-e^{-(4x^2+2I)}}}-1
 \end{align}
 which for positive $I$ and $x$ not too small (already $x\geq 1$ is good) is approximately zero. Therefore, we can approximate (for such values of $v$) $K(v,I)$ with $K(\infty,v)$. By using 
\begin{align}
 (1-x)^{-b} -1=\sum_{n=1}^\infty \frac {\Gamma(b+n)}{\Gamma(b)n!} x^n,
\end{align}
we get
\begin{align}
 K(\infty,v)=\frac 14 \sum_{n=1}^\infty \frac {\Gamma(n+\frac 12)}{\Gamma(\frac 12)n!} \sqrt {\frac \pi{n}} e^{-2In}.
\end{align}
Now, the Stirling formula gives a good approximation for the Gamma function for arguments larger than 1:
\begin{align}
 \Gamma(n+\frac 12)&\approx (n-\frac 12)^{n-\frac 12} e^{-n-\frac 12} \sqrt{2\pi(n-\frac 12)}\cr
 &\approx n^n e^{-n} \sqrt {2\pi n} (en)^{-\frac 12}\cr
 &\approx n! (en)^{-\frac 12}.
\end{align}
Therefore,
\begin{align}
 K(\infty,v)&\approx\frac 14 \sqrt {\frac \pi{e}} \sum_{n=1}^\infty \frac 1n e^{-2I_0 n}\cr
 &=-\frac 14 \sqrt {\frac \pi{e}} \log \left( 1-e^{-2I_0}\right).
\end{align}
Since $ \sqrt {\frac \pi{e}}\approx 1$, this justifies \eqref{ApproxI0Big}. However, the latter seems to become not so good when $I_0\to 0$ since it diverges. The reason is probably that the above derivation fails when $I_0$ is very small and $v$ also is small. In this case, $\partial_v K$ fails to be small, and the constant approximation is not good. In this situation, we conveniently rewrite
\begin{align}
 G(v,I)&=\frac 14 \int_0^{4v^2} \frac {ds}{\sqrt s} \frac 1{\sqrt {1-e^{-(s+2I)}}}\cr
 &=\frac 14 \int_{2I}^{4v^2+2I} \frac {ds}{\sqrt {s-2I}} \frac 1{\sqrt {1-e^{-s}}}.
\end{align}
Assuming that $I\leq 2v^2\ll 1$, we can approximate $\sqrt {1-e^{-s}}\approx \sqrt s$ so that
\begin{align}\notag
 G(v,I)&\approx \frac 14 \int_{1}^{1+4v^2/I} \frac {dt}{\sqrt {t^2-1}}\\ 
 &=\frac 14 \log \left( 1+\frac {4v^2}I +\sqrt {\frac {4v^2}I\Big( 2+\frac {4v^2}I \Big)} \right),
\end{align}
which is equivalent to
\begin{align}
 1+\frac {4v^2}I =\cosh (4G(v,I)).
\end{align}
Therefore, \eqref{equation} gives
\begin{align}
 u\approx p \sqrt {\frac {I_0}2} \sinh (4\pi \sqrt k L_r). \label{smallu}
\end{align}
This is expected to be the correct expression when $I_0$ is very small.\\
That this is the right behavior when $u\to 0$ can also be inferred from \eqref{equation}. Indeed, let us set $I_0\approx 2\lambda \bar u^2$, with $\bar u:=u(2\pi)/p$. Changing the variable $\tau$ into $\bar u p$, \eqref{equation} becomes
\begin{equation}
    2\pi \sqrt{k}L_r=\int_{0}^{1} \frac{\bar u dx}{\sqrt{1-\exp(-4\bar u^2 (x^2 +\lambda^2))}}. 
\end{equation}
Since $x^2+\lambda^2$ is bounded by $1+\lambda^2$, for $u$ very small, we can Taylor expand the exponential to first order, getting 
\begin{equation}
    2\pi \sqrt{k}L_r\approx \frac 12\int_{0}^{1} \frac{dx}{\sqrt{x^2 +\lambda^2}}=\frac 12 {\rm arcsinh} \frac 1{\lambda}. 
\end{equation}
Therefore,
\begin{equation}
    \frac 1\lambda=\sinh(4\pi \sqrt{k}L_r), 
\end{equation}
confirming \eqref{smallu}.

\section{The Jacobi theta constant $\theta_3(q)$}\label{App:Theta}
The third Jacobi theta function is 
\begin{align}
    \vartheta_3(z,q)=\sum_{n\in\mathbb Z} q^{n^2} e^{2i\pi z},
\end{align}
convergent for $|q|<1$. The associated theta constant is
\begin{align}
    \theta_3(q):=\vartheta_3(0,q)=\sum_{n\in\mathbb Z} q^{n^2}.
\end{align}
Let us introduce the function
\begin{align}
    \Theta(x):=\theta_3\left(e^{-\pi x}\right)=\sum_{n\in\mathbb Z} e^{-\pi x n^2},\qquad\text{Re}(x)>0.
\end{align}
By using the Poisson's resummation formula
\begin{eqnarray}
\sum_{n \in \mathbb Z} f(n)=\sum_{m \in \mathbb Z} \tilde f(m),
\end{eqnarray}
with 
\begin{eqnarray}
\tilde f (m)=\int_{-\infty}^\infty e^{-2i\pi m t} f(t),
\end{eqnarray}
to the function $\Theta$, with $f(y)=e^{-\pi x y^2}$ and $\tilde f(m)=e^{-\frac {\pi m^2}{x}}/\sqrt{x}$, we get 
\begin{eqnarray}\label{ThetaRel}
\Theta(1/x)=\sqrt x \ \Theta(x).
\end{eqnarray}
Notice that when Re$(x)\to+\infty$ then $\Theta(x)\to 1$, so, for Re$(x)\to 0^+$ we have
\begin{eqnarray}\label{ApproxTheta}
\Theta(x)=\frac {\Theta(1/x)}{\sqrt x}\approx \frac 1{\sqrt x}.
\end{eqnarray}
Finally, we notice that, for $Re(s)>0$,
\begin{align}
    \int_0^\infty (\Theta(x)-1) x^{\frac s2-1}dx=&2\sum_{n=1}^\infty \int_0^\infty e^{-\pi x n^2} x^{\frac s2-1}dx\cr
    =&\frac 2{\pi^{s/2}} \sum_{n=1}^\infty \frac 1{n^s}
    \int_0^\infty e^{-z} z^{\frac s2-1}dz\cr 
    =& \frac 2{\pi^{s/2}} \Gamma(s/2) \zeta(s).
\end{align}
 
\end{appendix}

\newpage

\bibliographystyle{plain}
\bibliography{MagSkyrme}

\end{document}